\documentclass[useAMS,usenatbib]{mn2e}

\usepackage{tabularx}
\usepackage{graphicx}
\usepackage{amsmath}
\usepackage{natbib}

\title[SCUBA-2 observations of WISE-selected Hot DOGs]{Submillimetre observations of WISE-selected high-redshift, luminous, dusty galaxies}

\author[Suzy F. Jones et al.]
{\parbox{\textwidth}{Suzy F. Jones,$^{1}$\thanks{E-mail: sfj8@le.ac.uk}
Andrew W. Blain,$^{1}$ 
Daniel Stern,$^{2}$
Roberto J. Assef,$^{3}$ 
Carrie R. Bridge,$^{4}$ 
Peter Eisenhardt,$^{2}$ 
Sara Petty,$^{5}$
Jingwen Wu,$^{6}$
Chao-Wei Tsai,$^{2}$
Roc Cutri,$^{7}$
Edward L. Wright$^{6}$
and Lin Yan$^{7}$
}\vspace{0.4cm}\\
\parbox{\textwidth}{$^{1}$University of Leicester, X-ray and Observational Astronomy Group (XROA), Department of Physics \& Astronomy, University Road, Leicester LE1 7RH, UK\\
$^2$Jet Propulsion Laboratory, California Institute of Technology, 4800 Oak Grove Dr., Pasadena, CA 91109, USA\\
$^3$N\'ucleo de Astronom\'ia de la Facultad de Ingenier\'ia, Universidad Diego Portales, Av. Ej\'ercito Libertador 441, Santiago, Chile\\
$^4$California Institute of Technology MS249-17, Pasadena, CA, 91125, USA\\
$^5$Virginia Polytechnic Institute \& State University, Department of Physics MC 0435, 850 West Campus Drive, Blacksburg, VA 24061, USA\\
$^6$Division of Physics \& Astronomy, University of California Los Angeles, Physics and Astronomy Building, 430 Portola Plaza, Los Angeles, CA 90095-1547, USA\\
$^7$Infrared Processing and Analysis Center, California Institute of Technology, MS 100-22, Pasadena, CA 91125, USA\\
}}

\begin{document}

\date{Submitted xx/xx/2014}

\pagerange{\pageref{firstpage}--\pageref{lastpage}} \pubyear{2014}

\maketitle

\label{firstpage}

\begin{abstract}

We present SCUBA-2 850\,$\mu$m submillimetre (submm) observations of the fields of 10 dusty, luminous galaxies at $z$ $\sim$ 1.7 - 4.6, detected at 12\,$\mu$m and/or 22\,$\mu$m by the WISE all-sky survey, but faint or undetected at 3.4\,$\mu$m and 4.6\,$\mu$m; dubbed hot, dust-obscured galaxies (Hot DOGs). The six detected targets all have total infrared luminosities greater than 10$^{13}$\,L$_{\odot}$, with one greater than 10$^{14}$\,L$_{\odot}$. Their spectral energy distributions (SEDs) are very blue from mid-infrared to submm wavelengths and not well fitted by standard AGN SED templates, without adding extra dust extinction to fit the WISE 3.4\,$\mu$m and 4.6\,$\mu$m data. The SCUBA-2 850\,$\mu$m observations confirm that the Hot DOGs have less cold and/or more warm dust emission than standard AGN templates, and limit an underlying extended spiral or ULIRG-type galaxy to contribute less than about 2$\%$ or 55$\%$ of the typical total Hot DOG IR luminosity, respectively. The two most distant and luminous targets have similar observed submm to mid-infrared ratios to the rest, and thus appear to have even hotter SEDs. The number of serendipitous submm galaxies (SMGs) detected in the 1.5-arcmin-radius SCUBA-2 850\,$\mu$m maps indicates there is a significant over-density of serendipitous sources around Hot DOGs. These submm observations confirm that the WISE-selected ultra-luminous galaxies have very blue mid-infrared to submm SEDs, suggesting that they contain very powerful AGN, and are apparently located in unusual arcmin-scale overdensities of very luminous dusty galaxies. 

\vspace{0.6cm}
\end{abstract} 


\section{Introduction}

Ultra-Luminous Infrared Galaxies (ULIRGs)\footnote{LIRGs, ULIRGs and HyLIRGs have characterising total infrared luminosities (8-1000\,$\mu$m) of $L_{8-1000\mu \textrm{m}} > 10^{11}$\,L$_\odot$, $L_{8-1000\mu \textrm{m}} > 10^{12}$\,L$_\odot$ and $L_{8-1000\mu \textrm{m}} > 10^{13}$\,L$_\odot$} were first discovered in the 1980's by the \textit{Infrared Astronomical Satellite} (\textit{IRAS}) \citep{houck84,soifer84}. More than 90\% of their luminosity is emitted in the infrared (IR) due to interstellar dust absorbing ultraviolet (UV) and optical emission produced by active galactic nuclei (AGN) and/or starbursts. The dust then re-emits thermally at longer wavelengths, from the near-IR to millimetre (mm) wavebands. 

ULIRGs evolve strongly with redshift, becoming more abundant with a surface density of several hundred per square degree at $z$ $\sim$ 1. Out to $z$ $\sim$ 1 the evolution rate of luminous dusty galaxies goes as $\sim$ (1 $+$ $z$)$^{4}$ \citep{blain99,floc05}. To $z$ $\ge$ 1 ULIRGs along with Luminous Infrared Galaxies (LIRGs),$^1$ account for 70 $\pm$ 15\% of cosmic star formation activity \citep{floch05,richards06}. Therefore, at the peak of the cosmic star formation rate, $z$ $\sim$ 2 - 3, ULIRGs contribute a significant amount to the total IR luminosity density \citep{smail97,genzel00,blain02,cowie02,chapman05,floch05,hopkins08,reddy08,magnelli09,elbaz11,casey12,magnelli12,melbourne12,lu13}. Studying the most extreme IR galaxies at this epoch, should provide a larger and more complete sample of different types of the most luminous AGN, to help in fully understanding the processes of formation and evolution of massive galaxies.   

A popular theory for the origin of ULIRGs is that major mergers between massive, gas-rich galaxies provide tidal torques that transport gas to the centre of the more massive galaxy \citep{barnes92,schweizer98,farrah01,veilleux02,hopkins06,hopkins08}. This influx of gas can induce rapid star formation and/or AGN fuelling \citep{barnes92,mihos96,hopkins08}: starburst activity dominates the luminosity at first, and then the embedded supermassive black hole (SMBH) grows to dominate. Feedback from the SMBH (radiation, winds and/or jets) and supernovae can expel gas and dust, terminating further star formation and for a short time leaving a visible optical quasar (QSO): finally, a passive massive elliptical galaxy is left behind \citep{sanders&mirabel96,hopkins06,hopkins08,farrah12,spoon13}. Observations of other dusty galaxy populations could be evidence of different stages of this merging galaxy theory. For example, submillimetre galaxies (SMGs) appear to be high redshift ULIRGs \citep{blain02,tacconi08} and dust-obscured galaxies (DOGs) have comparable star formation rates and IR luminosities to SMGs \citep{bussmann09,tyler09,melbourne12}. There could be an evolutionary connection between ULIRGs, SMGs, DOGs, QSOs and massive elliptical galaxies \citep{sanders88a,sanders88b}. 

Luminous, dusty active galaxies heated by AGN and/or recent starburst activity emit in the IR at wavelengths traced by the \textit{Wide-field Infrared Survey Explorer} (WISE) filters at 12\,$\mu$m (W3) and 22\,$\mu$m (W4) bands. \citet{eisenhardt12}, \citet{wu12} and Bridge et al. (2013, in prep.) have shown that WISE can find different classes of interesting, luminous, high-redshift, dusty galaxies. Based on WISE colours and flux cuts, a population has faint or undetectable flux densities in the 3.4\,$\mu$m (W1) and 4.6\,$\mu$m (W2) bands, while being well detected (signal-to-noise ratio (SNR) $>$ 5 in the All-Sky WISE Source Catalog\footnote{http://wise2.ipac.caltech.edu/docs/release/allsky/}) in the 12 and/or 22\,$\mu$m bands. These galaxies have been called ``W1W2-dropouts'' \citep{eisenhardt12} and Hot DOGs \citep{wu12}. 

In the major merger theory, the SMG population would represent an earlier, starburst-dominated phase of merging galaxies, and the luminous DOG population are the later, most luminous AGN-dominated phase of merging galaxies, and in energetic terms could easily become optically visible QSOs \citep{narayanan10}. The Hot DOGs presented in this paper would qualify the DOG selection criterion, F$_{24\mu \textrm{m}} >$ 0.3 mJy and R$- [24] >$ 14 (where R is the Vega magnitudes for optical R band and \textit{Spitzer} mid-IR 24\,$\mu$m \citep{dey08}) however, the Hot DOGs are more luminous, hotter and rarer than typical DOGs and could be extreme cases of DOGs or another stage of merging galaxies \citep{wu12}. To investigate this Hot DOG population, follow-up spectroscopy of more than 100 of them revealed that these galaxies are intrinsically very luminous, potentially putting them in the class of ULIRGs$^{1}$ and Hyper-Luminous Infrared Galaxies (HyLIRGs)$^{1}$ \citep{sanders&mirabel96,lonsdale06} (Bridge et al. 2013, in prep.; Eisenhardt et al. in prep.; Tsai et al. in prep.). They are frequently found in the redshift range 2 $<$ $z$ $<$ 3 \citep{eisenhardt12,wu12,bridge13}, and so far the highest redshift is $z = 4.59$ for W2246-0526 (Tsai et al. in prep.). Their SEDs show a mid-IR to far-IR colour that is too blue to be well fitted by many standard AGN templates, suggesting that they represent a short evolutionary phase of merging galaxies, where an AGN is fueling very rapidly inside a thick dust shroud, leading to very intense mid-IR but obscured emission and a hot SED, as proposed by (Wu et al. 2012; Bridge et al. 2013; Assef et al. in prep.). The Hot DOGs should show the impact of an AGN on the surrounding ISM at its very greatest. These WISE-selected sources are certainly not typical galaxies, but the processes taking place within them should be at work everywhere.

In this paper, James Clerk Maxwell Telescope (JCMT) Submillimetre Common-User Bolometer Array 2 (SCUBA-2) \citep{holland13} observations of 10 Hot DOGs are reported. These long wavelength measurements are needed to understand the cold dust properties and to calculate the total luminosity all the way from 8\,$\mu$m to 1000\,$\mu$m ($L_{8-1000\mu \textrm{m}}$). 
Section 2 describes the sample, along with the details of WISE and SCUBA-2 observations. 
Section 3 reports the SCUBA-2 results, and the SEDs and total IR luminosities ($L_{8-1000\mu \textrm{m}}$) of the Hot DOGs in comparison with other populations. Existing SED templates of well-studied objects are compared to find out the nature of the Hot DOGs, and their accuracy and a need for additional mid-IR extinction to fit the data is discussed. The submm to mid-IR ratios are discussed, to investigate if the Hot DOG SEDs are dominated by AGN emission or star formation. The luminosities of an underlying extended host galaxy component are calculated, in order to calculate the potential host galaxy contribution to the typical Hot DOG total IR luminosity, to see if the Hot DOGs are dominated by AGN or starburst activity. To see if there is an overdensity of SMGs in the SCUBA-2 fields, SMG number counts are compared to those in other submm surveys. 

Throughout this paper we assume a $\Lambda$CDM cosmology with H$_0$ = 71\,km\,s$^{-1}$Mpc$^{-1}$, $\Omega_{\rm{m}}$ = 0.27 and $\Omega_\Lambda$ = 0.73 \citep{hinshaw09}. WISE catalogue magnitudes are converted to flux densities using zero-point values on the Vega system of 306.7, 170.7, 29.04 and 8.284\,Jy for WISE 3.4, 4.6, 12 and 22\,$\mu$m wavelengths, respectively \citep{wright10}.

\section{Observations}

\subsection{WISE}

WISE was launched in December 2009 and surveyed the entire sky at wavelengths of 3.4, 4.6, 12 and 22\,$\mu$m \citep{wright10}. One of the primary science goals was to identify the most luminous galaxy in the observable universe, which can be accomplished due to WISE obtaining much greater sensitivity than previous all-sky IR survey missions. For example, \textit{IRAS} yielded catalogued source sensitivities of 0.5\,Jy at 12, 25 and 60\,$\mu$m and 1\,Jy at 100\,$\mu$m \citep{neugebauer84}. WISE achieved 5-$\sigma$ source sensitivities better than 0.054, 0.071, 0.73 and 5.0\,mJy and angular resolutions of 6.1, 6.4, 6.5 and 12.0\,arcsec in the W1 to W4 bands, respectively \citep{wright10,jarrett11}. 

\subsection{Target Selection}

The objects observed here are selected from the WISE All-Sky Source catalog$^2$, with IR magnitudes derived using point source profile-fitting \citep{cutri12}. The Hot DOG selection criteria are to have a faint or undetectable flux in W1 and W2, but a detectable flux (SNR $>$ 5) in W3 and/or or W4.  
The selected galaxies have W1 $>$ 17.4 mag and either W4 $<$ 7.7 mag and W2 $-$ W4 $>$ 8.2 or W3 $<$ 10.6 mag and W2 $-$ W3 $>$ 5.3 \citep{eisenhardt12}. The search was made greater than 30$^{\circ}$ away from the Galactic centre and 10$^{\circ}$ from the Galactic plane to avoid enhanced levels of saturation artifacts and stars. 

The number of WISE-selected Hot DOGs over the extragalactic sky, to this magnitude limit, is about 1000, which points to this population being extremely rare, and perhaps a transitional population (Eisenhardt et al. 2012; Assef et al. in prep.; Tsai et al. in prep.). The JCMT targets were selected because they had known spectroscopic redshifts (Eisenhardt et al. 2012, in prep.; Bridge et al. in prep.), could be observed in the A-semester (January to July) at the JCMT, and were queued to obtain \textit{Herschel} data (Bridge et al. in prep.; Tsai et al. in prep.). We selected 31 targets and obtained observations of 10 targets with SCUBA-2 that were chosen at random due to the vagaries of the queue observing system. Their WISE W4 band fluxes were selected to be among the greatest of the suitable sources, in the hope of increasing the chance that their 850\,$\mu$m flux would be bright enough to be detected or limits would be significant. Therefore, with any conclusions drawn it must be remembered that these Hot DOGs have been selected to be mid-IR bright.

The WISE flux densities presented in Table 1 are from the subsequent AllWISE Source Catalog\footnote{http://wise2.ipac.caltech.edu/docs/release/allwise/}, that has improved photometric sensitivity and accuracy, and improved astrometric precision compared to the WISE All-Sky Source Catalog.
 
\subsection{JCMT SCUBA-2}

Ten Hot DOGs were observed with SCUBA-2 on the 15-m JCMT atop Mauna Kea in Hawaii, primarily in May 2012 but also on other nights throughout the 12A semester, from January to July 2012. 

SCUBA-2 is a bolometer camera and has eight 32 x 40 pixel detector arrays each with a field of view of 2.4\,arcmin$^2$  \citep{holland13}. SCUBA-2 observes in the atmospheric windows at 450\,$\mu$m and 850\,$\mu$m. The diffraction-limited beams have full-width half maxima (FWHM) of approximately 7.5 and 14.5\,arcsec, respectively.

The optical depth at 225 GHz, $\tau$$_{225}$, during the observations was in the range of JCMT Band 2 conditions: 0.05 $<$ $\tau$$_{225}$ $<$ 0.08 \citep{dempsey13}. The corresponding opacities for each atmospheric window, 450\,$\mu$m and 850\,$\mu$m,  were 0.61 $<$ $\tau$$_{450}$ $<$ 1.18 and 0.24 $<$ $\tau$$_{850}$ $<$ 0.40. Therefore, we could not use any 450\,$\mu$m data because the atmospheric opacity was too great.

All observations were taken in the ``CV DAISY" mode that produces a 12-arcmin diameter map, with the deepest coverage in a central 3-arcmin diameter region \citep{holland13}. The target stays near the centre of the arrays and the telescope performs a pseudo-circular pattern with a radius of 250\,arcsec at a speed of 155\,arcsec\,s$^{-1}$. This mode is best for point-like sources and those smaller than 3-arcmin. Each scan was 30\,minutes long and four scans were made per target, totalling a exposure time of 120\,minutes per target. The typical 850\,$\mu$m noise achieved in these DAISY maps was 1.8\,mJy/beam, and the noise increases by $\sim$10$\%$ out to a radius of 1.5\,arcmin (Table 1). We have treated only this uniform central region of the SCUBA2 DAISY maps in this analysis.

Pointing checks were taken throughout the night. The calibration sources observed were Uranus, CRL 2688, CRL 618 and Mars. Calibrations were taken at the start and end of every night in the standard manner \citep{dempsey13}, and where consistent with the standard values.

\section{Results}

\subsection{Photometry} 

The maps were reduced with the STARLINK SubMillimeter User Reduction Facility (SMURF) data reduction package with the ``Blank Field" configuration suitable for low SNR point sources \citep{chapin13}. SMURF performs pre-processing steps to clean the data by modelling each of the contributions to the signal from each bolometer, flatfields and removes atmospheric emission, and finally regrids to produce a science-quality image. Using the STARLINK PIpeline for Combining and Analyzing Reduced Data (PICARD) package the maps were mosaiced with all four observations per target, beam-match filtered with a 15\,arcsec FWHM Gaussian and calibrated with the flux conversion factor (FCF) of 2.34\,Jy\,pW$^{-1}$\,arcsec$^{-2}$ (appropriate for aperture photometry) or 537\,Jy\,pW$^{-1}$\,beam$^{-1}$ (in order to measure absolute peak fluxes of discrete sources) that is appropriate for 850\,$\mu$m data \citep{dempsey13}.

\begin{figure*}
\includegraphics[width=17cm,height=11.5cm]{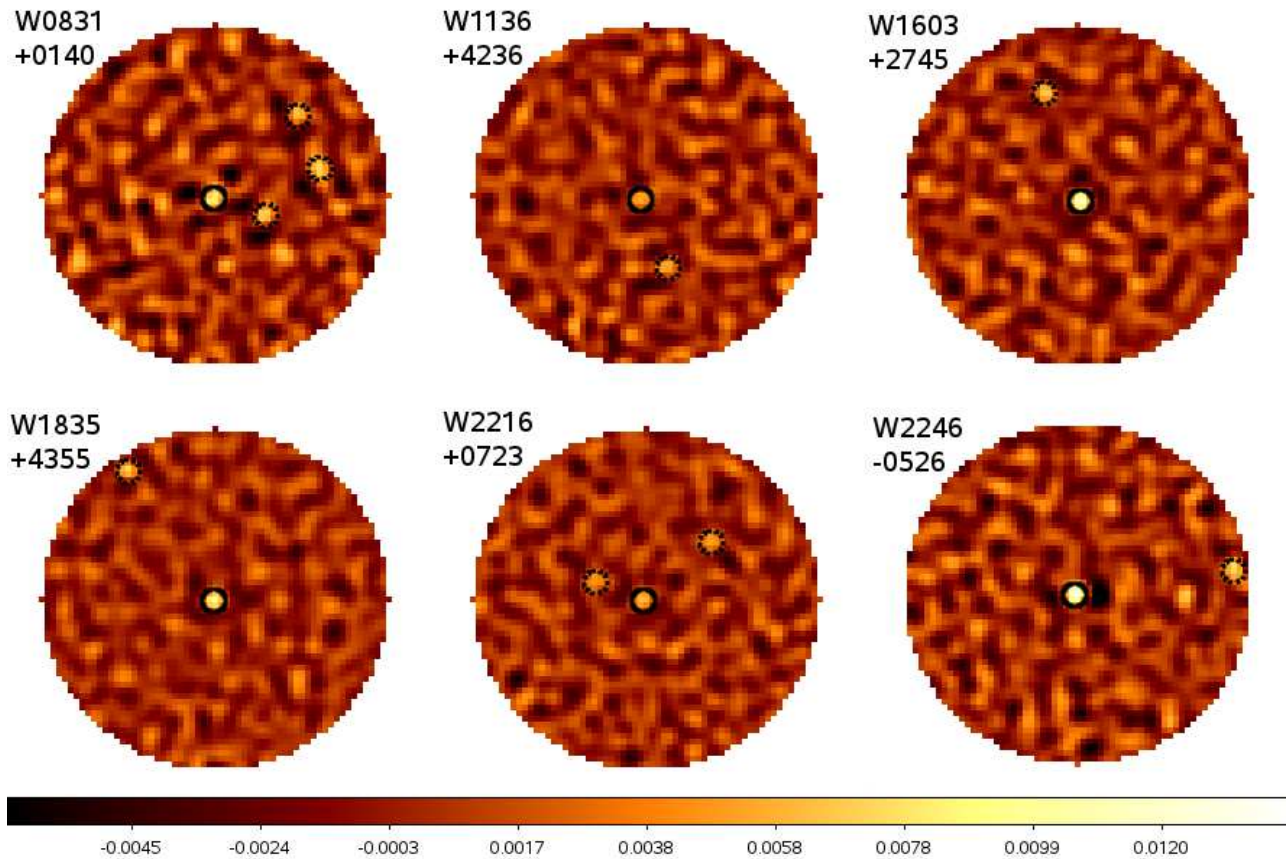}
\caption{SCUBA-2 850\,$\mu$m 1.5\,arcmin radius maps of the 6 detected targets; W0831$+$0140, W1136$+$4236, W1603$+$2745, W1835$+$4355, W2216$+$0723 and W2246$-$0526. The solid circles show the 15-arcmin beam-sized apertures centred on the WISE RA DEC of the targets. Serendipitous sources brighter than 3$\sigma$ and within 1.5\,arcmin radius of the WISE target are shown by the dotted 15-arcmin beam-sized circles. The colour flux bar at the bottom is in Jy. North is up, East is to the left.}
\label{det6}
\end{figure*}

\begin{figure*}
\includegraphics[width=11cm,height=11cm]{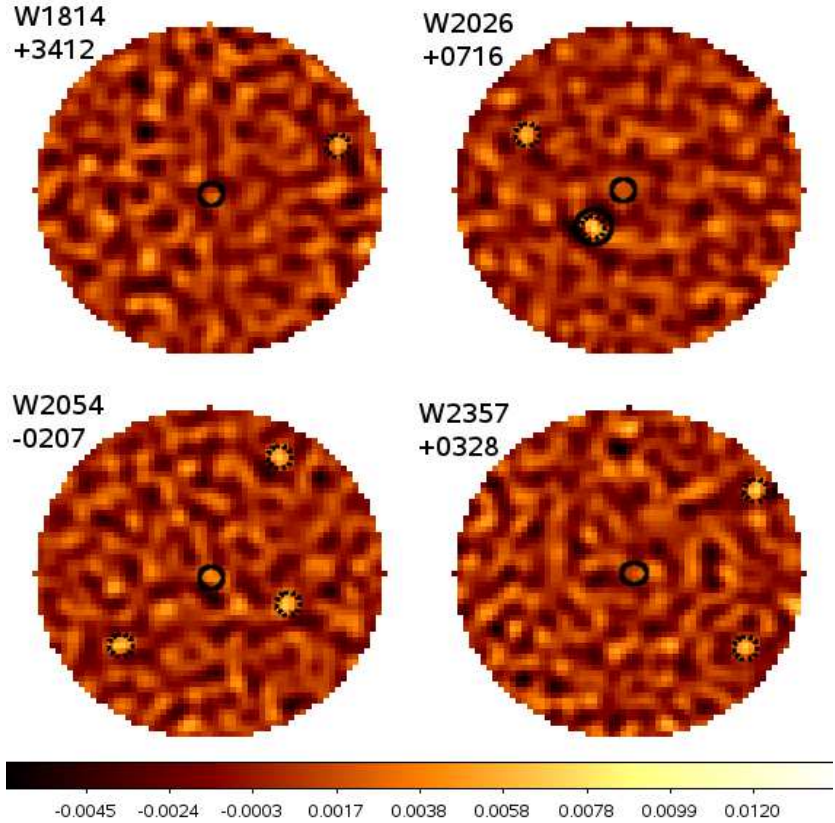}
\caption{SCUBA-2 850\,$\mu$m 1.5\,arcmin radius maps of the 4 undetected targets; W1814$+$3412, W2026$+$0716, W2054$+$0207 and W2357$+$0328. The solid circles show the 15-arcmin beam-sized apertures centred on the WISE RA DEC of the targets. Serendipitous sources brighter than 3$\sigma$ and within 1.5\,arcmin radius of the WISE target are shown by the dotted 15-arcmin beam-sized circles, and serendipitous source brighter than 4$\sigma$ is shown by the dotted 15-arcmin beam-sized circle surrounded by a solid black circle. The colour flux bar at the bottom is in Jy. North is up, East is to the left.}
\label{undet4}
\end{figure*}

The 850\,$\mu$m flux densities of the 10 Hot DOGs at their WISE positions and the noise level in the maps are presented in Table 1. Six Hot DOGs are detected at greater than 3$\sigma$ significance, while the other four targets had positive flux measurements at the WISE position with significances between 1.1$\sigma$ and 1.9$\sigma$. The flux density limits for all the targets were measured in an aperture diameter of 15\,arcsec, which is the same size as the FWHM of the telescope beam. This was an appropriate aperture size: for the detected sources, the aperture flux densities on this scale are consistent with the peak flux densities. Figure~\ref{det6} and Figure~\ref{undet4} show the sensitive 3-arcmin diameter SCUBA-2 850\,$\mu$m DAISY fields of the 6 detected Hot DOGs and the 4 undetected Hot DOGs, respectively. The typical error of the WISE position compared to the SCUBA-2 position of the detected targets was 1\,arcsec.

To test whether the positive flux density of the four targets with upper limits is likely to be real, random points were sampled from the maps, and the stacked average flux density was 0.0 $\pm$ 0.5\,mJy. This is consistent with the positive flux densities from the Hot DOGs with upper limits being due to fainter, undetected targets.

\begin{figure*}
\includegraphics[width=15cm]{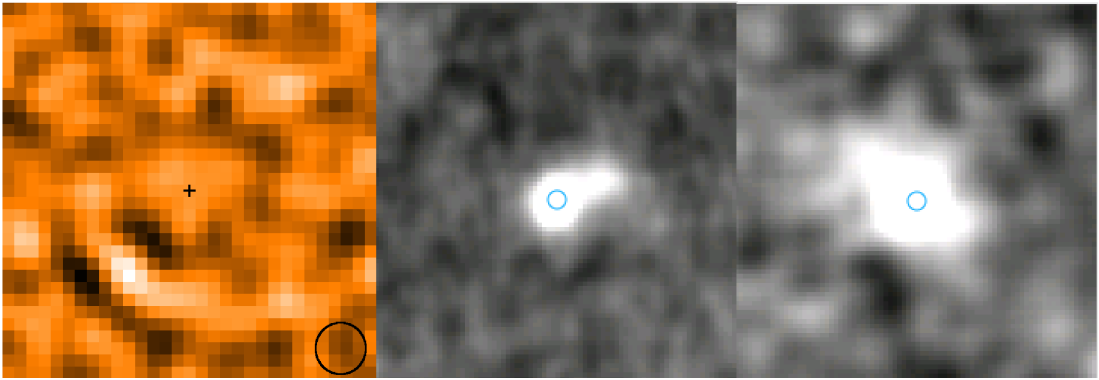}
\caption{Left: The 2\,arcmin $\times$ 2\,arcmin SCUBA-2 850\,$\mu$m map of W2026$+$0716. Increasing the aperture size out to 29\,arcsec in diameter increases the measured flux for this target, which could be due to multiple components of the target or to nearby, unrelated sources. The beam-size (15\,arcsec diameter) is represented by the black circle. Note the 4.4$\sigma$ serendipitous source $\sim$ 35\,arcsec south-east of the target. Centre and right: WISE W3 and W4 images, respectively, showing the same region of sky, with a circle showing the WISE determined position of the target. The target is not extended in the WISE images. North is up, and East is to the left.}
\label{w2026scuba2wise}
\end{figure*}

W2026$+$0716 is the only target whose 850\,$\mu$m flux increases when measured in a larger aperture (see Figure~\ref{w2026scuba2wise}). It has a 850\,$\mu$m flux density of 2.1 mJy with a 15-arcsec beam-sized aperture. However, when increasing the diameter to 29\,arcsec, the flux density increases to 7.3 mJy. The higher flux is likely because the target has multiple components on scales bigger than the SCUBA-2 beam. Multiple components have been seen in another Hot DOG, W1814$+$3412, where several objects on scales less than 10\,arcsec are apparent \citep{eisenhardt12}. Alternatively, there could be an unrelated source or sources. The WISE extended source flag is 0 for the W1 through W4 bands, which means that the detected source is not extended at WISE wavelengths (see Figure~\ref{w2026scuba2wise}). There are no obvious signs of a cluster of sources nearby in \textit{Spitzer} images (Wu et al. priv. comm.). However, \textit{Herschel} 160\,$\mu$m imaging shows a possible companion about 10\,arcsec away that could contribute enhanced flux in larger apertures in this source (Bridge et al. priv. comm.). Due to this uncertainty, the quoted 850\,$\mu$m flux density in Table 1 is the 15-arcsec beam-sized aperture flux density (2.1\,mJy).

\begin{figure}
\begin{centering}
\includegraphics[width=6cm,height=6cm]{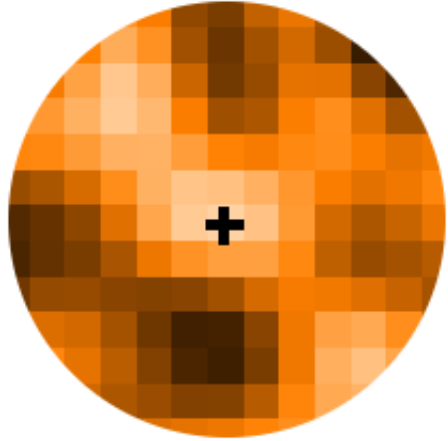}
\caption{SCUBA-2 850\,$\mu$m 48\,arcsec radius map showing the central region of the 4 undetected targets stacked together. The cross shows the central pixel. North is up, East is to the left.}
\label{stack}
\end{centering}
\end{figure}

When the images of the four undetected sources are stacked together into one image (Figure~\ref{stack}) and centred on the WISE-determined position of each target, the net flux is 7.8 $\pm$ 2.3\,mJy in the central 15\,arcsec region, a net detection of 3.4$\sigma$. The four undetected targets are consistent with being on average 2.5 times fainter than the six detected targets. To get deeper observations with SCUBA-2 would require several more hours of integration per target, beyond the existing 120\,minutes, but would not add much more value to this stacked result.

The flux densities presented in Table 1 can be compared with the results of \citet{wu12}, who observed 14 WISE-selected Hot DOGs with the Caltech Submillimeter Observatory (CSO) SubMillimeter High Angular Resolution Camera II (SHARC-II) at 350 to 850\,$\mu$m and 18 Hot DOGs with CSO Bolocam at 1.1 mm. Using a 3$\sigma$ threshold, \citet{wu12} found that nine out of 14 Hot DOGs were detected at 350\,$\mu$m and 6 of the 18 targets were detected at 1.1\,mm. Three sources from the Hot DOG sample in this paper were in common with \citet{wu12}; W1603+2745, W1814$+$3412 and W1835+4355. These CSO results are consistent with our SCUBA-2 observations at 850\,$\mu$m. The relative sensitivities of SHARC-II and Bolocam are such that we believe the SCUBA-2 detections and limits provide a substantial increase in our knowledge of the HotDOGs' submm properties. Furthermore, W1814$+$3412 which was detected at 350\,$\mu$m by \citet{wu12}, with upper limits reported at 450\,$\mu$m and 1100\,$\mu$m was also detected by the Institut de Radioastronomie Millimetrique (IRAM) Plateau de Bure interferometer in the 1.3\,mm band in 2013 (Blain et al. in prep.).

\subsection{SEDs}

\subsubsection{Short Wavelength SEDs}

\begin{figure*}
\includegraphics[width=170mm]{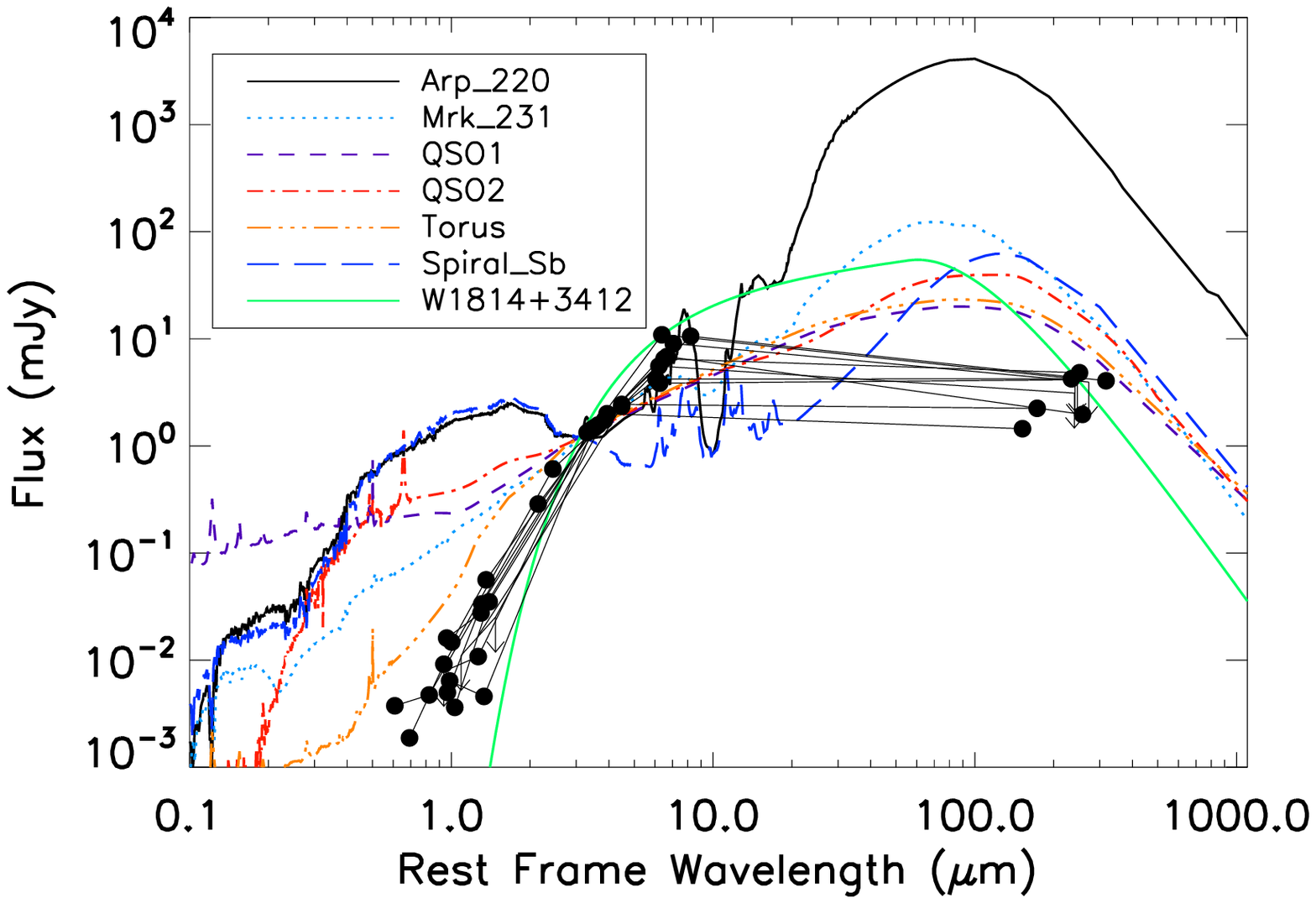}
\caption{SEDs of the 10 Hot DOGs including the 850\,$\mu$m SCUBA-2 data in rest-frame wavelengths with Arp 220, Mrk 231, QSO 1, QSO 2 and torus galaxy templates from \citet{polletta07} and W1814$+$3412 template from Blain et al. (in prep.) normalised at 3\,$\mu$m. Detections are represented by filled circles, while 2$\sigma$ upper limits are represented by arrows. The data points for the Hot DOGs are connected for clarity, and do not represent the true SED.}
\label{sedall}
\end{figure*}

The SEDs of the 10 SCUBA-2 Hot DOGs are shown in Figure~\ref{sedall}. The SEDs are normalised at rest-frame 3\,$\mu$m and shown at rest-frame wavelengths in order to compare to various galaxy SED templates \citep{polletta07} in order to try and understand the Hot DOGs nature. The Polletta galaxy templates are Arp 220 (starburst-dominated galaxy), Mrk 231 (heavily obscured AGN-starburst composite), QSO 1 and 2 (optically-selected QSOs of Type 1 and 2) and torus (type-2 heavily-obscured QSO: an accreting SMBH with a hot accretion disk surrounded by dust and Compton-thick gas in a toroidal structure \citep{krolik88}. The Hot DOG SEDs are broadly similar. They have a steep red power-law IR (1-5\,$\mu$m) section with a potential mid-IR peak from hot dust emission, a mid-IR to submm section that appears to be flatter, i.e. less peaked, than the Polletta AGN templates, turning over to a Rayleigh-Jeans spectrum longwards of 200\,$\mu$m from the coolest dust emission. The mid-IR to submm section is consistent with being ``flat-topped'' (in f$_{\nu}$), as suggested by \citet{wu12}, and consistent with the 850\,$\mu$m data presented in this paper, and consistent to \textit{Herschel} results from the most luminous Hot DOGs (Tsai et al. in prep.). However, \textit{Herschel} data of Lyman-alpha blobs (LABs) \citep{bridge13} that have similar WISE colours to the Hot DOGs in this paper show a far-IR peak in the SED. Further discussion of this point and the presentation of \textit{Herschel} follow-up of WISE Hot DOGs will be presented by Bridge et al. in prep. and Tsai et al. in prep.

\begin{figure}
\includegraphics[width=84mm]{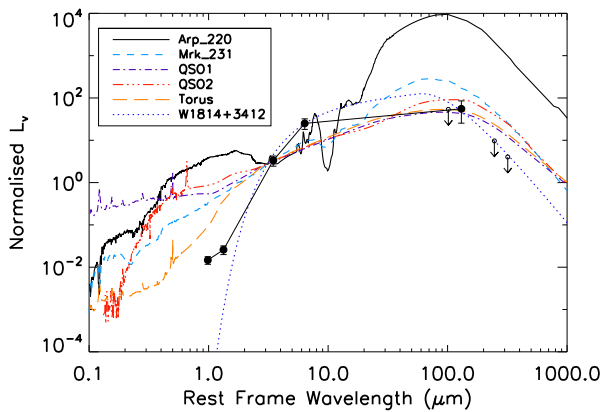}
\caption{SED of W1814$+$3412 at $z$=2.452, including the 850\,$\mu$m SCUBA-2 data with Polletta galaxy templates Arp 220, Mrk 231, QSO 1, QSO 2 and torus \citep{polletta07} and W1814$+$3412 template (Blain et al. in prep.) normalised at rest-frame 3\,$\mu$m. CSO SHARC-II 350\,$\mu$m and 450\,$\mu$m and CSO Bolocam 1100\,$\mu$m data points from \citet{wu12} are included. Detections are represented by filled circles, while 2$\sigma$ upper limits are represented by arrows.}
\label{w1814_temp}
\end{figure}

A better fitting SED model for these Hot DOGs is also shown in Figure~\ref{sedall} and Figure~\ref{w1814_temp}. The W1814$+$3412 template shown is entirely empirical, and assumes a single-temperature dust spectrum representing the minimum dust temperature present (53 $\pm$ 5\,K), with an emissivity index of $\beta$ = 1.5 at longer wavelengths, smoothly interpolated to a power-law spectrum instead of a Wien law at shorter wavelengths, with an opacity factor imposed at the shortest mid-IR wavelengths, to match the WISE data, corresponding to a finite total luminosity $L_{8-1000\mu \textrm{m}}$ = 4.6 $\times$ 10$^{13}$\,L$_\odot$ for W1814$+$3412. It is constrained by \textit{Herschel} data from \citet{wu12} and IRAM data from Blain et al. (in prep.), and so unsurprisingly provides a better fit than the \citet{polletta07} templates that pre-date these observations.

The SEDs are not well-fitted by any of the templates; the closest fitting template is the single torus template, although extra dust extinction is required to fit the W1 and W2 data. Between the Polletta torus template and the mean SED of the 850\,$\mu$m detected targets, both normalised at 3\,$\mu$m, the extra dust extinction required at rest-frame 1\,$\mu$m is 1.6\,mag. Converting to a $V$-band extinction implies an extra dust extinction $A_{\rm{V}}$ $\ge 6.8$\,mag. \citet{eisenhardt12} found significant obscuration in the SED of W1814$+$3412 with a dust extinction value of $A_{\rm{V}} = 48 \pm 4$ in the rest-frame from optical SED fitting. The gas column density $N_{\rm{H}}$ can be estimated by applying a standard ``gas-to-extinction" equation $N_{\rm{H}} \approx 2 A_{\rm{V}} \times 10^{22}$\,mag$^{-1}$\,cm$^{-2}$ from \citet{maiolino01}, which estimates the extra $N_{\rm{H}}$ needed is $10^{23}$\,cm$^{-2}$. The Polletta torus template was modelled on a heavily obscured type-2 QSO, with a rest-frame $N_{\rm{H}}$ of $2.14^{+0.54}_{-1.34} \times 10^{24}$\,cm$^{-2}$ \citep{polletta06}. The extra $N_{\rm{H}}$ can be added to the Polletta torus template $N_{\rm{H}}$ to estimate the total $N_{\rm{H}}$ of the Hot DOGs to be $\sim$ 2.3 $\times 10^{24}$\,cm$^{-2}$, implying a Compton-thick AGN \citep{osterbrock88,madau94,comastri95,maiolino95,risaliti99,piconcelli03,treister05}. This is consistent with \citet{stern14}, who observed three Hot DOGs, including W1814$+$3412 in common with this paper, with NuSTAR and XMM-Newton, and found that the three targets have gas column densities $N_{\rm{H}}$ $\ge$ $10^{24}$\,cm$^{-2}$, which implies the targets are Compton-thick AGNs.

\subsubsection{Long Wavelength SEDs}

Normalised to the WISE data at rest-frame 3\,$\mu$m that lies within the WISE rest-frame wavelength range for all of our Hot DOGs, the SCUBA-2 data shows that the Hot DOGs have less submm emission than the Polletta torus template, with an average flux difference factor of 5 between the data and template, and a range of 2-8 factor, for the detected targets, and an average flux difference factor of 7 between the data and template, and a range of 6-8, for the undetected targets (limits plus 2$\sigma$). 

\begin{figure}
\includegraphics[width=84mm]{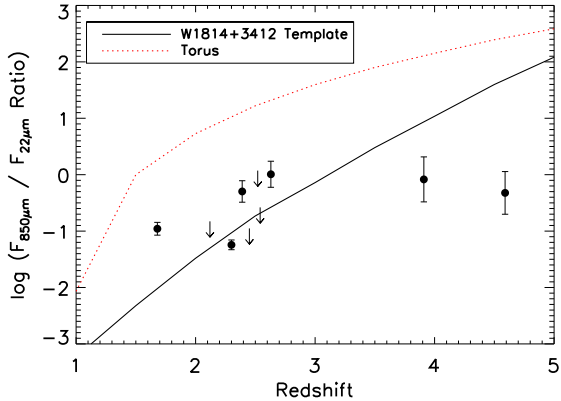}
\caption{The submm to mid-IR ratio (F$_{850 \mu \textrm{m}}$ / F$_{22 \mu \textrm{m}}$) of the 10 Hot DOGs. The solid line shows the W1814$+$3412 template (Blain et al. in prep.) and the dotted line shows Polletta torus template \citep{polletta07} as a function of redshift. Detections are represented by filled circles, while 2$\sigma$ upper limits are represented by arrows.}
\label{85022}
\end{figure}

The submm to mid-IR ratio (F$_{850 \mu \textrm{m}}$ / F$_{22 \mu \textrm{m}}$) of the 10 targets in the observed frame are listed in Table 1. The weighted average F$_{850 \mu \textrm{m}}$ / F$_{22 \mu \textrm{m}}$ of the six detected targets is 0.6 $\pm$ 0.1, where the error is the weighted standard error. Figure~\ref{85022} shows these ratios in the observed sample, and for the W1814$+$3412 and Polletta AGN torus templates, as a function of redshift. The subsequent empirical W1814$+$3412 template provides a better fit to the targets, as expected. Most of the Hot DOGs appear to lie near the W1814$+$3412 template but show no clear sign of the expected K-correction with redshift of the templates: in particular the two highest redshift Hot DOGs (W0831$+$0140 and W2246$-$0526) lie beneath both templates with relatively faint submm fluxes, and similar observed submm to mid-IR ratios of the other lower redshift Hot DOGs. They are also the most luminous targets, with infrared luminosities, $L_{8-1000\mu \textrm{m}}$ $>$ 10$^{14}$\,L$_{\odot}$ (see section 3.3), which suggests that they have hotter effective dust temperatures compared with the rest of the sample. Another distant Hot DOG, W0410-0913 at $z$ = 3.592, was reported by \citet{wu12}. It has a submm to mid-IR ratio of 3.2 $\pm$ 0.8 (log(0.5 $\pm$ 0.1)), and lies close to the W1814$+$3412 template.

The Hot DOGs are all consistent with high dust temperatures inferred from the submm/WISE data \citep{wu12}. The W1814$+$3412 template (Figure~\ref{w1814_temp}) has a temperature of 53 $\pm$ 5\,K for the coolest contribution to the SED. Estimates of the temperatures for Hot DOGs and WISE-selected LABs have included 60-120K \citep{wu12} and 40-90K \citep{bridge13}. This is greater than other SMGs and DOGs, which have typical temperatures of 25 - 40\,K \citep{chapman05,coppin08,magnelli12,melbourne12}. The definitions of SED shape and temperature contributions can be complex, but the rest-frame peak of the Hot DOG SEDs occurs bluewards of other galaxy classes.

\subsection{Luminosities}

Total IR luminosities are calculated for each target using a minimal power-law interpolation between the WISE and SCUBA-2 data points, and also by using the W1814$+$3412 template with each targets' WISE and SCUBA-2 data. 

A conservative lower limit to the total IR luminosities of the galaxies were estimated by connecting all the WISE and SCUBA-2 data points with power-laws and then integrating, without extrapolating beyond the range of the data in wavelength. This luminosity is a conservative estimate because any strong peak in the SED would not be included on the power-law interpolation. The resulting $L_{8\mu \textrm{m}-\textrm{SCUBA2}}$ values for the 10 Hot DOGs are presented in Table 2. The six detected targets have values that range from (1.0 $\pm$ 1.8) $\times$ 10$^{13}$\,L$_{\odot}$ to (1.3 $\pm$ 2.9) $\times$ 10$^{14}$\,L$_{\odot}$, with one having $L_{8\mu \textrm{m}-\textrm{SCUBA2}}$ $\ge$ 10$^{14}$\,L$_{\odot}$. This classifies them all as HyLIRGs. 

The $L_{8-1000\mu \textrm{m}}$ are also found by fitting the W1814$+$3412 template from Blain et al. (in prep.) with the targets' SCUBA-2 data. The W1814$+$3412 template was used because it is based on data from WISE, \textit{Herschel} and IRAM of the Hot DOG W1814$+$3412 that are near the peak of the SED. The six detected targets assuming the W1814$+$3412 template, yield luminosities from (2.7 $\pm$ 1.6) $\times$ 10$^{13}$\,L$_{\odot}$ to (6.4 $\pm$ 2.4) $\times$ 10$^{14}$\,L$_{\odot}$, the factor of $\sim$4 difference between the two methods for calculating the luminosities is due to the systematic uncertainties of the SED. 

Our total IR luminosity of W1835$+$4355 ($L_{8\mu \textrm{m}-\textrm{SCUBA2}}$  = 4.0 $\pm$ 4.1 $\times$ 10$^{13}$\,L$_{\odot}$), the source also observed by CSO, is consistent with $L_{8-1000\mu \textrm{m}}$  = 6.5 $\times$ 10$^{13}$\,L$_{\odot}$ reported in \citet{wu12}. The derived luminosity of the W1814$+$3412 template from Blain et al. (in prep.) was $L_{8-1000\mu \textrm{m}}$ = 4.6 $\times$ 10$^{13}$\,L$_\odot$ and compared with the luminosity calculated using the WISE and SCUBA-2 data was $L_{8\mu \textrm{m}-\textrm{SCUBA2}}$ $<$ 2.0 $\times$ 10$^{13}$\,L$_\odot$, is slightly higher due to the W1814$+$3412 template included CSO SHARC-II data, which is nearer the peak of the SED.

Exceptionally bright galaxies can often be found to be gravitational lensed, for example \citet{eisenhardt96,williams96,solomon05,vieira10,negrello10,bussmann13}. However, the Hot DOG luminosities are thought to be intrinsic and not due to gravitational lensing: high-resolution imaging programmes (Bridge et al. in prep.; Petty et al. in prep) from \textit{Hubble Space Telescope} (\textit{HST}) and ground-based telescopes of a subset of Hot DOGs show no obvious lensed structures \citep{wu14}. Resolved near-IR \textit{HST} observations show the population to have a range of morphologies from clumpy and extended to point-like (Bridge et al. in prep.; Petty et al. in prep.). This suggests that the Hot DOGs are indeed amongst the most intrinsically luminous galaxies in the universe (Eisenhardt et al. 2012; Tsai et al. in prep.).

The four undetected Hot DOGs each have $L_{8\mu \textrm{m}-\textrm{SCUBA2}}$ $\le$ 10$^{13}$\,L$_{\odot}$ (Table 2). The stacked 850\,$\mu$m flux density (7.8 $\pm$ 2.3\,mJy) of the four targets with 850\,$\mu$m upper limits was used with the W1814$+$3412 template to estimate the luminosity $L_{8-1000\mu \textrm{m}}$ = (9.3 $\pm$ 4.7) $\times$ 10$^{13}$\,L$_\odot$, which is consistent with a HyLIRG. A higher luminosity could be found if there are \textit{Herschel} detections of a significant peak in the far-IR (Bridge et al. in prep.; Tsai et al. in prep.).

We also use the SCUBA-2 data to limit the luminosity of an underlying extended galaxy. A spiral (Sb) galaxy template and a warmer ULIRG-type (Arp 220) template were fitted to account for all of the SCUBA-2 850\,$\mu$m flux density, and then by integrating under the Sb or Arp 220 template, the maximum total luminosity of this template SED can be estimated. This approach assumes that an underlying extended dusty galaxy, disconnected from the mid-IR emission, accounts for all of the measured SCUBA-2 flux. This extended emission can be assumed to all be due to star-formation rather than an AGN. An Sb host galaxy template cannot exceed $\sim$ 2$\%$ of the  inferred Hot DOG luminosity from 8-1000\,$\mu$m. This would give a Sb luminosity of 1.3 $\times$ 10$^{12}$\,L$_{\odot}$; 22 times more luminous than the Milky Way (6 $\times$ 10$^{10}$\,L$_{\odot}$), with an equivalent star formation rate (SFR) of $\sim$ 30\,M$_\odot$\,yr$^{-1}$, which is lower than the UV-derived SFR of $\sim$ 300\,M$_\odot$\,yr$^{-1}$ derived for W1814$+$3412 \citep{eisenhardt12}. An Arp 220 ULIRG template can account for the 850\,$\mu$m data, if the host galaxy has a luminosity of 2.9 $\times$ 10$^{13}$\,L$_{\odot}$, $\sim$ 55$\%$ of the inferred Hot DOG luminosity, with an equivalent SFR of $\sim$ 450\,M$_\odot$\,yr$^{-1}$. This emphasises that the full Hot DOG SED from 8-1000\,$\mu$m has a small contribution from cold far-IR dust and is dominated by hot dust and mid-IR emission.

\subsection{Clustering}

There is significant evidence from previous studies that the galaxy density in the environments of high-redshift far-IR and mid-IR luminous galaxies and SMGs appears to be above average \citep{scott02,blain04,borys04,scott06,farrah06,gilli07,chapman09,cooray10,hickox12}. Clustering of SMGs could be evidence of massive dark matter halos associated with the SMGs at high-redshift. To investigate if there is clustering of SMGs in the Hot DOG fields, the serendipitous sources number counts will be compared with the number counts from two different blank-field submm surveys. To provide another way to test SMG clustering in the Hot DOG fields, 1.5-arcmin-radius circles will be places at random and centred on SMG detections in a blank-field submm survey.

Seventeen serendipitous 850\,$\mu$m sources were detected at greater than 3$\sigma$ in the 10 SCUBA-2 maps, and one source was detected at greater than 4$\sigma$; see Table 3. The total area surveyed is 71\,arcmin$^2$, or about 1500 SCUBA-2 850\,$\mu$m beams. Figure~\ref{det6} and Figure~\ref{undet4} show the location of detected serendipitous sources in the SCUBA-2 fields of all the detected and undetected Hot DOGs, respectively. 

There are 4$\pm$2 negative peaks in the 10 maps at above the same 3$\sigma$ threshold (see Table 3), consistent with the 2$\pm$1 3$\sigma$ negative peaks expected from Gaussian noise.

To see if there is evidence for an over-density of SMGs in the 10 SCUBA-2 Hot DOG fields, the number of serendipitous sources can be compared with the results of field submm surveys. In the LESS survey, \citet{weiss09} detected 126 SMGs in a uniform area of 1260\,arcmin$^2$ with a noise level of 1.2\,mJy at 870\,$\mu$m. They also found evidence for an angular two-point clustering signal on angular scales smaller than 1\,arcmin. There are 101 LESS sources brighter than our average 3$\sigma$ flux density limit of 5.3\,mJy, which implies 5.7 serendipitous sources would be expected in our 10 SCUBA-2 fields; however, we find 15. This indicates a relative overdensity of SMGs in our Hot DOG fields by a factor of 2.6 $\pm$ 0.7. The noise level range of our maps is 1.5-2.1 mJy beam$^{-1}$. In order to check the effect of our range of sensitivity in each field we also compare the number of SMGs at our highest noise level (2.1 mJy beam$^{-1}$) with to the LESS survey. The number of LESS sources brighter than our greatest 3$\sigma$ flux density limit of 6.3\,mJy is 60 SMGs, which implies that 3.4 serendipitous sources would be expected in our 10 SCUBA-2 fields. However, we find 9 and thus a relative overdensity of SMGs by a factor of 2.7 $\pm$ 1.0. The overdensity using the highest noise level is consistent with that using the average noise level; therefore, the difference in the 10 map noise levels appears not to have a large effect on the overdensity factor.

A complementary way to test whether there is an overdensity of SMGs near the Hot DOG targets is to place 1.5-arcmin-radius circles at random locations in the LESS field and count the number of sources from the catalogue source positions that would have been detected in our survey, taking into account the differences in depth, by employing a flux density limit of 5.3\,mJy. The 1.5-arcmin-radius circles were chosen because the SCUBA-2 maps were 1.5 arcmin in radius. In 10 sets of 10 randomly selected 1.5-arcmin-radius circles within LESS, we found 7 $\pm$ 3 LESS sources brighter than 5.3\,mJy, to mimic our SCUBA-2 images in this surveyed field. The total number of 1.5-arcmin-radius fields available within LESS is $\sim$ 200. There is thus a hint of evidence for a relative over-density of SMGs around Hot DOGs by a factor of 2.1 $\pm$ 1.0 as compared with this blank-field.

A third way to test the over density of SMGs in the SCUBA-2 fields is to compare the number of LESS sources brighter than 5.3\,mJy within 1.5-arcmin-radius circles centred on LESS-detected sources. In 101 available positions, there are 18 not counting the LESS sources on which each 1.5-arcmin-radius field was centred. This suggests that in 10 SCUBA-2 fields centred on LESS detections there would be only 1.8 serendipitous sources detected; however, we find 15 centred on WISE-selected targets, potentially giving a Hot DOG to SMG companion overdensity factor of order 8.

We can repeat this approach in another submm blank-field survey. \citet{casey13} used SCUBA-2 to observe the COSMOS field over a uniform area of 394\,arcmin$^2$ at a noise level of 0.80\,mJy at 850\,$\mu$m and detected 99 SMGs brighter than 3.6$\sigma$. There are 18 COSMOS sources brighter than our average detection threshold of 5.3\,mJy (3$\sigma$), which would imply 6.3 serendipitous sources expected in the 10 SCUBA-2 Hot DOG fields. We find 15, implying a relative overdensity of SMGs by a factor of 2.4 $\pm$ 0.7, which is consistent with 2.6 $\pm$ 0.7 from LESS. The number of sources brighter than our greatest 3$\sigma$ flux density limit of 6.3\,mJy is 18 SMGs, which implies that 3.2 serendipitous sources would be expected in our 10 SCUBA-2 fields. However, we find 9 and thus a relative overdensity of SMGs by a factor of 2.8 $\pm$ 1.1, which is consistent with 2.7 $\pm$ 1.0 from LESS. Again, the variation in noise levels of the 10 Hot DOG field does not appear to have a large effect on the overdensity factor. These results are consistent with the LESS survey. Random 1.5-arcmin-radius circles were not placed in the COSMOS field due to the small size of the field, with only $\sim$ 50 1.5-arcmin-radius fields available. 

The six detected Hot DOG targets and the four undetected Hot DOG targets have similar numbers of serendipitous sources, see Figure~\ref{det6} and Figure~\ref{undet4}. The overdensity of SMGs around detected and undetected Hot DOGs appear to be comparable.

\begin{figure}
\includegraphics[width=84mm]{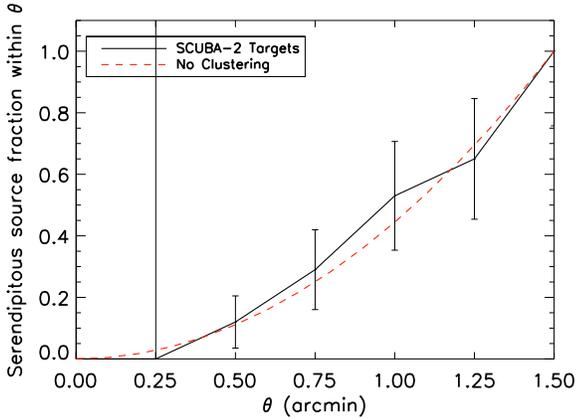}
\caption{The fraction of the total number of serendipitous sources in each field within different radii of the WISE targets. The solid line shows the fields of the Hot DOGs. The dashed-dotted line shows the expected number of serendipitous sources if they are randomly located with no clustering. The beam size of SCUBA-2 at 850\,$\mu$m is 14.5\,arcsec; serendipitous sources cannot be detected within the beam.}
\label{clustering_his}
\end{figure}

Figure~\ref{clustering_his} shows the fraction of the total number of serendipitous sources for the 10 SCUBA-2 maps found within 0.25, 0.5, 0.75, 1.0, 1.25 and 1.5\,arcmin from the WISE target. The expected fraction of the total number of serendipitous sources with no angular clustering is also plotted on Figure~\ref{clustering_his}. There is no hint of angular clustering of serendipitous sources around the Hot DOGs on these scales, despite the greater average density of submm sources in the WISE fields as compared with blank-field surveys. Clustering of SMGs on larger scales could be expected because there is tentative evidence of clustering from previous submm studies on scales up to $\sim$8\,arcmin \citep{scoville00,blain04,greve04,farrah06,ivison07,weiss09,cooray10,scott10,hickox12}. Nevertheless, the lack of a clear two-point correlation signal is interesting, because SMG clustering observations can constrain of the nature of the host halos around SMGs \citep{cooray10}.

\section{Discussion}

The above results obtained for the Hot DOGs will be discussed by comparing their SEDs and luminosities with other galaxy populations. Next the Hot DOGs environments are investigated by comparing the serendipitous source number counts to other submm surveys. 

The Hot DOG SEDs in Figure~\ref{sedall} show a blue mid-IR to submm colour. A Compton-thick AGN torus template would fit the Hot DOG SEDs if extra mid-IR extinction of $A_{\rm{V}}$ $\ge$ 6.8\,mag is included. These results can be compared with the SEDs of ULIRGs and LIRGs which have large amounts of obscuring material around the AGN and/or starburst activity, and have estimated dust extinctions between $A_{\rm{V}}$ $\simeq$ 5 and 50\,mag \citep{genzel98}. Comparing the SED of Arp 220 (a starburst-dominated ULIRG) with the Hot DOG SEDs in Figure~\ref{sedall}, reveals that the Hot DOGs appear to have more mid-IR dust extinction than the exceptionally mid-IR red Arp 220. This leads to the conclusion that the Hot DOGs have extremely large amounts of absorption in the AGN torus, and/or host galaxy \citep{goulding12}, and could have even larger amounts of obscuring material than typical ULIRGs and LIRGs.

The SCUBA-2 observations show that the Hot DOGs have relatively less submm emission than other galaxy SED templates: the detected Hot DOG targets' SCUBA-2 flux density is on average 5 times fainter than the Polletta torus template. This leads to the suggestion that the Hot DOGs have less cold dust in the host galaxy and/or on the outer edge of the torus, and hence the torus could be denser, smaller and hotter than in the template. Alternatively, less submm emission could be due to an excess of mid-IR emission from the AGN as compared with the torus template \citep{wu12}. The median 850\,$\mu$m flux density of SMGs, 5.7 $\pm$ 3.0\,mJy \citep{chapman05}, is comparable with these Hot DOGs with a median 850\,$\mu$m flux density of 5.4 $\pm$ 1.8\,mJy. Since SMGs and Hot DOGs have similar redshifts ($z \sim 2$), this might suggest comparable cold dust properties. However, to address the degree of similarity between SMGs and Hot DOGs in the far-IR will require knowledge of the far-IR colours of Hot DOGs with data from \textit{Herschel} (Bridge et al. in prep.; Tsai et al. in prep.).

The luminosities of all six detected targets (with a mean luminosity of $L_{8\mu \textrm{m}-\textrm{SCUBA2}}$  = 5.3 $\times$ 10$^{13}$\,L$_{\odot}$ and a median luminosity of $L_{8\mu \textrm{m}-\textrm{SCUBA2}}$  = 3.6 $\times$ 10$^{13}$\,L$_{\odot}$) are greater than those of typical SMGs, which have $L_{8-1000\mu \textrm{m}}$  = 8.5 $\times$ 10$^{12}$\,L$_{\odot}$ \citep{chapman05,kovacs06}, and DOGs, which have a mean luminosity $L_{8-1000\mu \textrm{m}}$  = 9 $\times$ 10$^{12}$\,L$_{\odot}$ \citep{melbourne12}. This is in agreement with \citet{wu12} who found a sample mean $L_{8-1000\mu \textrm{m}}$ = 6.1 $\times$ 10$^{13}$\,L$_{\odot}$ for Hot DOGs. However, the targets in this paper and \citet{wu12} could be biased towards being the mid-IR brightest and rarest galaxies, because they were selected on the grounds of their bright mid-IR flux. It is certainly inevitable that deeper mid-IR samples will include DOGs and SMGs; however, our current observed sample of 10, which has a range of W4 fluxes (6.19 to 7.66 mag, or 7.2 to 27.7\,mJy), due to the various selection cuts involved, certainly shows no SMG/ULIRG type SEDs. Eight WISE-selected LABs \citep{bridge13} were also found to be ultra-luminous galaxies from \textit{Herschel} data ($L_{8-1000\mu \textrm{m}}$ = 2.3 $\times$ 10$^{13}$\,L$_{\odot}$), and included a wider range of mid-IR fluxes with no colour cut, again indicating very luminous mid-IR properties of galaxies with extremely red WISE colours. 

Comparing number counts of the serendipitous sources in the 10 Hot DOG fields with other submm surveys, implies there is an overdensity of SMGs in the 10 SCUBA-2 fields by factor of $\sim$2-3. This is consistent with finding Hot DOGs in potentially overdense environments. \citet{umehata14} observed the protocluster SSA22 field with the Astronomical Thermal Emission Camera (AzTEC) on the Atacama Submillimeter Telescope Experiment (ASTE), at 1.1-mm to a depth of 0.7 - 1.3 mJy beam$^{-1}$, and found 10 SMGs are correlated with $z = 3.1$ Lyman-alpha emitters (LAE)s in the protocluster, which suggests that SMGs are formed in dense environments. Our SMG overdensity around Hot DOGs could indicate that Hot DOGs signpost protocluster regions.

These Hot DOGs appear to be very powerful AGN that have more mid-IR emission and mid-IR opacity than AGN in standard galaxy templates. Therefore, the Hot DOGs might be experiencing the most powerful feedback possible and could be an AGN-dominated short evolutionary phase of merging galaxies, and appear to reside in intriguing arcmin-scale overdensities of very luminous, dusty sources. 

\section{Summary}

The results from SCUBA-2 850\,$\mu$m observations of 10 WISE-selected, high-redshift, luminous, dusty Hot DOGs are:

\begin{itemize}
\item The 10 Hot DOGs have SEDs that are not well fitted by the current AGN templates (see Figure~\ref{sedall}). The best fitting single Polletta torus template \citep{polletta07} needs extra dust extinction to fit the Hot DOG SEDs with extra $A_{\rm{V}}$ $\ge$ 6.8\,mag, which could be due to more screening from the host galaxy and/or AGN torus. The $N_{\rm{H}}$ was estimated to be $\sim$ 2.3 $\times$ 10$^{24}$\,cm$^{-2}$, which is Compton-thick.

\item The Hot DOGs have a lower ratio of cold to hot dust than the Polletta torus template, which could be because there is less cold dust in the host galaxy, and/or the outer AGN torus in the Hot DOGs are smaller. Alternatively there could be more intense mid-IR emission from the inner regions \citep{wu12}. \textit{Herschel} observations near the peak of the SED should soon provide more information.

\item Despite being observed over a wide redshift range, the Hot DOGs show uniform submm to mid-IR ratios. The highest redshift, most luminous targets, could thus have hotter dust temperatures than assumed in the templates. However, the number of targets involved is currently only modest and the selection of the targets is sensitive to redshift, owing to very red intense WISE colours.

\item The six SCUBA-2 detected Hot DOGs have very high IR luminosities, $L_{8\mu \textrm{m}-\textrm{SCUBA2}}$ $\ge 10^{13}$\,L$_{\odot}$: they are HyLIRGs. These are conservative values as any pronounced peak of the SED would increase these further and could be missed without \textit{Herschel} data. The stacked IR luminosity, $L_{8-1000\mu \textrm{m}}$ = (9.3 $\pm$ 4.7) $\times$ 10$^{13}$\,L$_\odot$, of the four undetected targets is consistent with being a HyLIRG. With no obvious signatures of gravitational lensing known, Hot DOGs are amongst the most luminous galaxies.

\item The luminosity of an underlying extended star-forming galaxy cannot exceed a luminosity $\sim$ 2$\%$ (for a cool spiral galaxy template) or $\sim$ 55$\%$ (for a warmer ULIRG-like galaxy template) as compared with the typical Hot DOG luminosity, respectively. Our SCUBA-2 observations confirm that Hot DOGs are a mid-IR dominated population.

\item When comparing the submm galaxy counts of the 10 1.5-arcmin-radius SCUBA-2 maps observed here to blank-field surveys, there is an over-density of SMGs on this scale by a factor 3, but no evidence for any angular clustering within these fields. 

\item The next step to understand these Hot DOGs is with more \textit{Herschel} observations to accurately define the peak of the SED and increase the sample size, presented in future papers (Bridge et al. in prep.; Tsai et al. in prep.). Another subpopulation of galaxies selected with similar WISE colours but also selected to be radio bright, that could be AGN quenching star formation by radio jet feedback at the highest rate of AGN fueling. A larger number of these targets have been observed with SCUBA-2 and Atacama Large Millimeter/submillimeter Array (ALMA), and will be presented in future papers (Jones et al. in prep.; Lonsdale et al. in prep.). These SCUBA-2 observations provide a comparable density analysis.

\end{itemize}

\section{Acknowledgements}

The authors would like to thank the anonymous referee for his/her comments and suggestions, which have greatly improved this paper.

S. F. Jones gratefully acknowledges support from the University of Leicester Physics \& Astronomy Department. R. J. Assef was supported by Gemini-CONICYT grant number 32120009. This publication makes use of data products from the \textit{Wide-field Infrared Survey Explorer}, which is a joint project of the University of California, Los Angeles, and the Jet Propulsion Laboratory/California Institute of Technology, funded by the National Aeronautics and Space Administration.

The James Clerk Maxwell Telescope is operated by the Joint Astronomy Centre on behalf of the Science and Technology Facilities Council of the United Kingdom, the Netherlands Organisation for Scientific Research, and the National Research Council of Canada. Additional funds for the construction of SCUBA-2 were provided by the Canada Foundation for Innovation. The program ID under which the data were obtained was M12AU010.

\bibliographystyle{mn2e}
\bibliography{ref}

\begin{thebibliography}{86}
\expandafter\ifx\csname natexlab\endcsname\relax\def\natexlab#1{#1}\fi

\bibitem[{{Barnes} \& {Hernquist}(1992)}]{barnes92}
{Barnes} J.~E., {Hernquist} L., 1992, \araa, 30, 705

\bibitem[{{Blain} {et~al}\mbox{.}(2004){Blain}, {Chapman}, {Smail}, \&
  {Ivison}}]{blain04}
{Blain} A.~W., {Chapman} S.~C., {Smail} I., {Ivison} R., 2004, \apj, 611, 725

\bibitem[{{Blain} {et~al}\mbox{.}(1999){Blain}, {Smail}, {Ivison}, \&
  {Kneib}}]{blain99}
{Blain} A.~W., {Smail} I., {Ivison} R.~J., {Kneib} J.-P., 1999, \mnras, 302,
  632

\bibitem[{{Blain} {et~al}\mbox{.}(2002){Blain}, {Smail}, {Ivison}, {Kneib}, \&
  {Frayer}}]{blain02}
{Blain} A.~W., {Smail} I., {Ivison} R.~J., {Kneib} J.-P., {Frayer} D.~T., 2002,
  \physrep, 369, 111

\bibitem[{{Borys} {et~al}\mbox{.}(2004){Borys}, {Scott}, {Chapman}, {Halpern},
  {Nandra}, \& {Pope}}]{borys04}
{Borys} C., {Scott} D., {Chapman} S., {Halpern} M., {Nandra} K., {Pope} A.,
  2004, \mnras, 355, 485

\bibitem[{{Bridge} {et~al}\mbox{.}(2013){Bridge}, {Blain}, {Borys}, {Petty},
  {Benford}, {Eisenhardt}, {Farrah}, {Griffith}, {Jarrett}, {Lonsdale},
  {Stanford}, {Stern}, {Tsai}, {Wright}, \& {Wu}}]{bridge13}
{Bridge} C.~R. {et~al.}, 2013, \apj, 769, 91

\bibitem[{{Bussmann} {et~al}\mbox{.}(2009){Bussmann}, {Dey}, {Borys}, {Desai},
  {Jannuzi}, {Le Floc'h}, {Melbourne}, {Sheth}, \& {Soifer}}]{bussmann09}
{Bussmann} R.~S. {et~al.}, 2009, \apj, 705, 184

\bibitem[{{Bussmann} {et~al}\mbox{.}(2013){Bussmann}, {P{\'e}rez-Fournon},
  {Amber}, {Calanog}, {Gurwell}, {Dannerbauer}, {De Bernardis}, {Fu}, {Harris},
  {Krips}, {Lapi}, {Maiolino}, {Omont}, {Riechers}, {Wardlow}, {Baker},
  {Birkinshaw}, {Bock}, {Bourne}, {Clements}, {Cooray}, {De Zotti}, {Dunne},
  {Dye}, {Eales}, {Farrah}, {Gavazzi}, {Gonz{\'a}lez Nuevo}, {Hopwood}, {Ibar},
  {Ivison}, {Laporte}, {Maddox}, {Mart{\'{\i}}nez-Navajas}, {Michalowski},
  {Negrello}, {Oliver}, {Roseboom}, {Scott}, {Serjeant}, {Smith}, {Smith},
  {Streblyanska}, {Valiante}, {van der Werf}, {Verma}, {Vieira}, {Wang}, \&
  {Wilner}}]{bussmann13}
{Bussmann} R.~S. {et~al.}, 2013, \apj, 779, 25

\bibitem[{{Casey} {et~al}\mbox{.}(2012){Casey}, {Berta}, {B{\'e}thermin},
  {Bock}, {Bridge}, {Budynkiewicz}, {Burgarella}, {Chapin}, {Chapman},
  {Clements}, {Conley}, {Conselice}, {Cooray}, {Farrah}, {Hatziminaoglou},
  {Ivison}, {le Floc'h}, {Lutz}, {Magdis}, {Magnelli}, {Oliver}, {Page},
  {Pozzi}, {Rigopoulou}, {Riguccini}, {Roseboom}, {Sanders}, {Scott},
  {Seymour}, {Valtchanov}, {Vieira}, {Viero}, \& {Wardlow}}]{casey12}
{Casey} C.~M. {et~al.}, 2012, \apj, 761, 140

\bibitem[{{Casey} {et~al}\mbox{.}(2013){Casey}, {Chen}, {Cowie}, {Barger},
  {Capak}, {Ilbert}, {Koss}, {Lee}, {Le Floc'h}, {Sanders}, \&
  {Williams}}]{casey13}
{Casey} C.~M. {et~al.}, 2013, \mnras, 436, 1919

\bibitem[{{Chapin} {et~al}\mbox{.}(2013){Chapin}, {Berry}, {Gibb}, {Jenness},
  {Scott}, {Tilanus}, {Economou}, \& {Holland}}]{chapin13}
{Chapin} E.~L., {Berry} D.~S., {Gibb} A.~G., {Jenness} T., {Scott} D.,
  {Tilanus} R.~P.~J., {Economou} F., {Holland} W.~S., 2013, \mnras, 430, 2545

\bibitem[{{Chapman} {et~al}\mbox{.}(2009){Chapman}, {Blain}, {Ibata}, {Ivison},
  {Smail}, \& {Morrison}}]{chapman09}
{Chapman} S.~C., {Blain} A., {Ibata} R., {Ivison} R.~J., {Smail} I., {Morrison}
  G., 2009, \apj, 691, 560

\bibitem[{{Chapman} {et~al}\mbox{.}(2005){Chapman}, {Blain}, {Smail}, \&
  {Ivison}}]{chapman05}
{Chapman} S.~C., {Blain} A.~W., {Smail} I., {Ivison} R.~J., 2005, \apj, 622,
  772

\bibitem[{{Comastri} {et~al}\mbox{.}(1995){Comastri}, {Setti}, {Zamorani}, \&
  {Hasinger}}]{comastri95}
{Comastri} A., {Setti} G., {Zamorani} G., {Hasinger} G., 1995, \aap, 296, 1

\bibitem[{{Cooray} {et~al}\mbox{.}(2010){Cooray}, {Amblard}, {Wang},
  {Arumugam}, {Auld}, {Aussel}, {Babbedge}, {Blain}, {Bock}, {Boselli}, {Buat},
  {Burgarella}, {Castro-Rodriguez}, {Cava}, {Chanial}, {Clements}, {Conley},
  {Conversi}, {Dowell}, {Dwek}, {Eales}, {Elbaz}, {Farrah}, {Fox},
  {Franceschini}, {Gear}, {Glenn}, {Griffin}, {Halpern}, {Hatziminaoglou},
  {Ibar}, {Isaak}, {Ivison}, {Khostovan}, {Lagache}, {Levenson}, {Lu},
  {Madden}, {Maffei}, {Mainetti}, {Marchetti}, {Marsden}, {Mitchell-Wynne},
  {Mortier}, {Nguyen}, {O'Halloran}, {Oliver}, {Omont}, {Page}, {Panuzzo},
  {Papageorgiou}, {Pearson}, {Perez Fournon}, {Pohlen}, {Rawlings}, {Raymond},
  {Rigopoulou}, {Rizzo}, {Roseboom}, {Rowan-Robinson}, {Schulz}, {Scott},
  {Serra}, {Seymour}, {Shupe}, {Smith}, {Stevens}, {Symeonidis}, {Trichas},
  {Tugwell}, {Vaccari}, {Valtchanov}, {Vieira}, {Vigroux}, {Ward}, {Wright},
  {Xu}, \& {Zemcov}}]{cooray10}
{Cooray} A. {et~al.}, 2010, \aap, 518, L22

\bibitem[{{Coppin} {et~al}\mbox{.}(2008){Coppin}, {Halpern}, {Scott}, {Borys},
  {Dunlop}, {Dunne}, {Ivison}, {Wagg}, {Aretxaga}, {Battistelli}, {Benson},
  {Blain}, {Chapman}, {Clements}, {Dye}, {Farrah}, {Hughes}, {Jenness}, {van
  Kampen}, {Lacey}, {Mortier}, {Pope}, {Priddey}, {Serjeant}, {Smail},
  {Stevens}, \& {Vaccari}}]{coppin08}
{Coppin} K. {et~al.}, 2008, \mnras, 384, 1597

\bibitem[{{Cowie} {et~al}\mbox{.}(2002){Cowie}, {Barger}, \& {Kneib}}]{cowie02}
{Cowie} L.~L., {Barger} A.~J., {Kneib} J.-P., 2002, \aj, 123, 2197

\bibitem[{{Cutri} {et~al}\mbox{.}(2012){Cutri} {et~al.}}]{cutri12}
{Cutri} R.~M., {et~al.}, 2012, VizieR Online Data Catalog, 2311, 0

\bibitem[{{Dempsey} {et~al}\mbox{.}(2013){Dempsey}, {Friberg}, {Jenness},
  {Tilanus}, {Thomas}, {Holland}, {Bintley}, {Berry}, {Chapin}, {Chrysostomou},
  {Davis}, {Gibb}, {Parsons}, \& {Robson}}]{dempsey13}
{Dempsey} J.~T. {et~al.}, 2013, \mnras, 430, 2534

\bibitem[{{Dey} {et~al}\mbox{.}(2008){Dey}, {Soifer}, {Desai}, {Brand}, {Le
  Floc'h}, {Brown}, {Jannuzi}, {Armus}, {Bussmann}, {Brodwin}, {Bian},
  {Eisenhardt}, {Higdon}, {Weedman}, \& {Willner}}]{dey08}
{Dey} A. {et~al.}, 2008, \apj, 677, 943

\bibitem[{{Eisenhardt} {et~al}\mbox{.}(1996){Eisenhardt}, {Armus}, {Hogg},
  {Soifer}, {Neugebauer}, \& {Werner}}]{eisenhardt96}
{Eisenhardt} P.~R., {Armus} L., {Hogg} D.~W., {Soifer} B.~T., {Neugebauer} G.,
  {Werner} M.~W., 1996, \apj, 461, 72

\bibitem[{{Eisenhardt} {et~al}\mbox{.}(2012){Eisenhardt}, {Wu}, {Tsai},
  {Assef}, {Benford}, {Blain}, {Bridge}, {Condon}, {Cushing}, {Cutri}, {Evans},
  {Gelino}, {Griffith}, {Grillmair}, {Jarrett}, {Lonsdale}, {Masci}, {Mason},
  {Petty}, {Sayers}, {Stanford}, {Stern}, {Wright}, \& {Yan}}]{eisenhardt12}
{Eisenhardt} P.~R.~M. {et~al.}, 2012, \apj, 755, 173

\bibitem[{{Elbaz} {et~al}\mbox{.}(2011){Elbaz}, {Dickinson}, {Hwang},
  {D{\'{\i}}az-Santos}, {Magdis}, {Magnelli}, {Le Borgne}, {Galliano},
  {Pannella}, {Chanial}, {Armus}, {Charmandaris}, {Daddi}, {Aussel}, {Popesso},
  {Kartaltepe}, {Altieri}, {Valtchanov}, {Coia}, {Dannerbauer}, {Dasyra},
  {Leiton}, {Mazzarella}, {Alexander}, {Buat}, {Burgarella}, {Chary}, {Gilli},
  {Ivison}, {Juneau}, {Le Floc'h}, {Lutz}, {Morrison}, {Mullaney}, {Murphy},
  {Pope}, {Scott}, {Brodwin}, {Calzetti}, {Cesarsky}, {Charlot}, {Dole},
  {Eisenhardt}, {Ferguson}, {F{\"o}rster Schreiber}, {Frayer}, {Giavalisco},
  {Huynh}, {Koekemoer}, {Papovich}, {Reddy}, {Surace}, {Teplitz}, {Yun}, \&
  {Wilson}}]{elbaz11}
{Elbaz} D. {et~al.}, 2011, \aap, 533, A119

\bibitem[{{Farrah} {et~al}\mbox{.}(2006){Farrah}, {Lonsdale}, {Borys}, {Fang},
  {Waddington}, {Oliver}, {Rowan-Robinson}, {Babbedge}, {Shupe}, {Polletta},
  {Smith}, \& {Surace}}]{farrah06}
{Farrah} D. {et~al.}, 2006, \apjl, 641, L17

\bibitem[{{Farrah} {et~al}\mbox{.}(2001){Farrah}, {Rowan-Robinson}, {Oliver},
  {Serjeant}, {Borne}, {Lawrence}, {Lucas}, {Bushouse}, \& {Colina}}]{farrah01}
{Farrah} D. {et~al.}, 2001, \mnras, 326, 1333

\bibitem[{{Farrah} {et~al}\mbox{.}(2012){Farrah}, {Urrutia}, {Lacy},
  {Efstathiou}, {Afonso}, {Coppin}, {Hall}, {Lonsdale}, {Jarrett}, {Bridge},
  {Borys}, \& {Petty}}]{farrah12}
{Farrah} D. {et~al.}, 2012, \apj, 745, 178

\bibitem[{{Genzel} \& {Cesarsky}(2000)}]{genzel00}
{Genzel} R., {Cesarsky} C.~J., 2000, \araa, 38, 761

\bibitem[{{Genzel} {et~al}\mbox{.}(1998){Genzel}, {Lutz}, {Sturm}, {Egami},
  {Kunze}, {Moorwood}, {Rigopoulou}, {Spoon}, {Sternberg}, {Tacconi-Garman},
  {Tacconi}, \& {Thatte}}]{genzel98}
{Genzel} R. {et~al.}, 1998, \apj, 498, 579

\bibitem[{{Gilli} {et~al}\mbox{.}(2007){Gilli}, {Comastri}, \&
  {Hasinger}}]{gilli07}
{Gilli} R., {Comastri} A., {Hasinger} G., 2007, \aap, 463, 79

\bibitem[{{Goulding} {et~al}\mbox{.}(2012){Goulding}, {Alexander}, {Bauer},
  {Forman}, {Hickox}, {Jones}, {Mullaney}, \& {Trichas}}]{goulding12}
{Goulding} A.~D., {Alexander} D.~M., {Bauer} F.~E., {Forman} W.~R., {Hickox}
  R.~C., {Jones} C., {Mullaney} J.~R., {Trichas} M., 2012, \apj, 755, 5

\bibitem[{{Greve} {et~al}\mbox{.}(2004){Greve}, {Ivison}, {Bertoldi},
  {Stevens}, {Dunlop}, {Lutz}, \& {Carilli}}]{greve04}
{Greve} T.~R., {Ivison} R.~J., {Bertoldi} F., {Stevens} J.~A., {Dunlop} J.~S.,
  {Lutz} D., {Carilli} C.~L., 2004, \mnras, 354, 779

\bibitem[{{Hickox} {et~al}\mbox{.}(2012){Hickox}, {Wardlow}, {Smail}, {Myers},
  {Alexander}, {Swinbank}, {Danielson}, {Stott}, {Chapman}, {Coppin}, {Dunlop},
  {Gawiser}, {Lutz}, {van der Werf}, \& {Wei{\ss}}}]{hickox12}
{Hickox} R.~C. {et~al.}, 2012, \mnras, 421, 284

\bibitem[{{Hinshaw} {et~al}\mbox{.}(2009){Hinshaw}, {Weiland}, {Hill},
  {Odegard}, {Larson}, {Bennett}, {Dunkley}, {Gold}, {Greason}, {Jarosik},
  {Komatsu}, {Nolta}, {Page}, {Spergel}, {Wollack}, {Halpern}, {Kogut},
  {Limon}, {Meyer}, {Tucker}, \& {Wright}}]{hinshaw09}
{Hinshaw} G. {et~al.}, 2009, \apjs, 180, 225

\bibitem[{{Holland} {et~al}\mbox{.}(2013){Holland}, {Bintley}, {Chapin},
  {Chrysostomou}, {Davis}, {Dempsey}, {Duncan}, {Fich}, {Friberg}, {Halpern},
  {Irwin}, {Jenness}, {Kelly}, {MacIntosh}, {Robson}, {Scott}, {Ade},
  {Atad-Ettedgui}, {Berry}, {Craig}, {Gao}, {Gibb}, {Hilton}, {Hollister},
  {Kycia}, {Lunney}, {McGregor}, {Montgomery}, {Parkes}, {Tilanus}, {Ullom},
  {Walther}, {Walton}, {Woodcraft}, {Amiri}, {Atkinson}, {Burger}, {Chuter},
  {Coulson}, {Doriese}, {Dunare}, {Economou}, {Niemack}, {Parsons},
  {Reintsema}, {Sibthorpe}, {Smail}, {Sudiwala}, \& {Thomas}}]{holland13}
{Holland} W.~S. {et~al.}, 2013, \mnras, 430, 2513

\bibitem[{{Hopkins} {et~al}\mbox{.}(2006){Hopkins}, {Hernquist}, {Cox}, {Di
  Matteo}, {Robertson}, \& {Springel}}]{hopkins06}
{Hopkins} P.~F., {Hernquist} L., {Cox} T.~J., {Di Matteo} T., {Robertson} B.,
  {Springel} V., 2006, \apjs, 163, 1

\bibitem[{{Hopkins} {et~al}\mbox{.}(2008){Hopkins}, {Hernquist}, {Cox}, \&
  {Kere{\v s}}}]{hopkins08}
{Hopkins} P.~F., {Hernquist} L., {Cox} T.~J., {Kere{\v s}} D., 2008, \apjs,
  175, 356

\bibitem[{{Houck} {et~al}\mbox{.}(1984){Houck}, {Soifer}, {Neugebauer},
  {Beichman}, {Aumann}, {Clegg}, {Gillett}, {Habing}, {Hauser}, {Low}, {Miley},
  {Rowan-Robinson}, \& {Walker}}]{houck84}
{Houck} J.~R. {et~al.}, 1984, \apjl, 278, L63

\bibitem[{{Ivison} {et~al}\mbox{.}(2007){Ivison}, {Greve}, {Dunlop}, {Peacock},
  {Egami}, {Smail}, {Ibar}, {van Kampen}, {Aretxaga}, {Babbedge}, {Biggs},
  {Blain}, {Chapman}, {Clements}, {Coppin}, {Farrah}, {Halpern}, {Hughes},
  {Jarvis}, {Jenness}, {Jones}, {Mortier}, {Oliver}, {Papovich},
  {P{\'e}rez-Gonz{\'a}lez}, {Pope}, {Rawlings}, {Rieke}, {Rowan-Robinson},
  {Savage}, {Scott}, {Seigar}, {Serjeant}, {Simpson}, {Stevens}, {Vaccari},
  {Wagg}, \& {Willott}}]{ivison07}
{Ivison} R.~J. {et~al.}, 2007, \mnras, 380, 199

\bibitem[{{Jarrett} {et~al}\mbox{.}(2011){Jarrett}, {Cohen}, {Masci}, {Wright},
  {Stern}, {Benford}, {Blain}, {Carey}, {Cutri}, {Eisenhardt}, {Lonsdale},
  {Mainzer}, {Marsh}, {Padgett}, {Petty}, {Ressler}, {Skrutskie}, {Stanford},
  {Surace}, {Tsai}, {Wheelock}, \& {Yan}}]{jarrett11}
{Jarrett} T.~H. {et~al.}, 2011, \apj, 735, 112

\bibitem[{{Kov{\'a}cs} {et~al}\mbox{.}(2006){Kov{\'a}cs}, {Chapman}, {Dowell},
  {Blain}, {Ivison}, {Smail}, \& {Phillips}}]{kovacs06}
{Kov{\'a}cs} A., {Chapman} S.~C., {Dowell} C.~D., {Blain} A.~W., {Ivison}
  R.~J., {Smail} I., {Phillips} T.~G., 2006, \apj, 650, 592

\bibitem[{{Krolik} \& {Begelman}(1988)}]{krolik88}
{Krolik} J.~H., {Begelman} M.~C., 1988, \apj, 329, 702

\bibitem[{{Le Floc'h} {et~al}\mbox{.}(2005{\natexlab{a}}){Le Floc'h},
  {Papovich}, {Dole}, {Bell}, {Lagache}, {Rieke}, {Egami},
  {P{\'e}rez-Gonz{\'a}lez}, {Alonso-Herrero}, {Rieke}, {Blaylock},
  {Engelbracht}, {Gordon}, {Hines}, {Misselt}, {Morrison}, \& {Mould}}]{floc05}
{Le Floc'h} E. {et~al.}, 2005{\natexlab{a}}, \apj, 632, 169

\bibitem[{{Le Floc'h} {et~al}\mbox{.}(2005{\natexlab{b}}){Le Floc'h},
  {Papovich}, {Dole}, {Bell}, {Lagache}, {Rieke}, {Egami},
  {P{\'e}rez-Gonz{\'a}lez}, {Alonso-Herrero}, {Rieke}, {Blaylock},
  {Engelbracht}, {Gordon}, {Hines}, {Misselt}, {Morrison}, \&
  {Mould}}]{floch05}
{Le Floc'h} E. {et~al.}, 2005{\natexlab{b}}, \apj, 632, 169

\bibitem[{{Lonsdale} {et~al}\mbox{.}(2006){Lonsdale}, {Farrah}, \&
  {Smith}}]{lonsdale06}
{Lonsdale} C.~J., {Farrah} D., {Smith} H.~E., 2006, {Ultraluminous Infrared
  Galaxies}, Springer Verlag, p. 285

\bibitem[{{Lu} {et~al}\mbox{.}(2013){Lu}, {Zhao}, {Xu}, {Gao}, \& {GOALS FTS
  Team}}]{lu13}
{Lu} N., {Zhao} Y., {Xu} C.~K., {Gao} Y., {GOALS FTS Team}, 2013, in IAU
  Symposium, Vol. 292, IAU Symposium, {Wong} T., {Ott} J., eds., pp. 249--249

\bibitem[{{Madau} {et~al}\mbox{.}(1994){Madau}, {Ghisellini}, \&
  {Fabian}}]{madau94}
{Madau} P., {Ghisellini} G., {Fabian} A.~C., 1994, \mnras, 270, L17

\bibitem[{{Magnelli} {et~al}\mbox{.}(2009){Magnelli}, {Elbaz}, {Chary},
  {Dickinson}, {Le Borgne}, {Frayer}, \& {Willmer}}]{magnelli09}
{Magnelli} B., {Elbaz} D., {Chary} R.~R., {Dickinson} M., {Le Borgne} D.,
  {Frayer} D.~T., {Willmer} C.~N.~A., 2009, \aap, 496, 57

\bibitem[{{Magnelli} {et~al}\mbox{.}(2012){Magnelli}, {Lutz}, {Santini},
  {Saintonge}, {Berta}, {Albrecht}, {Altieri}, {Andreani}, {Aussel},
  {Bertoldi}, {B{\'e}thermin}, {Bongiovanni}, {Capak}, {Chapman}, {Cepa},
  {Cimatti}, {Cooray}, {Daddi}, {Danielson}, {Dannerbauer}, {Dunlop}, {Elbaz},
  {Farrah}, {F{\"o}rster Schreiber}, {Genzel}, {Hwang}, {Ibar}, {Ivison}, {Le
  Floc'h}, {Magdis}, {Maiolino}, {Nordon}, {Oliver}, {P{\'e}rez Garc{\'{\i}}a},
  {Poglitsch}, {Popesso}, {Pozzi}, {Riguccini}, {Rodighiero}, {Rosario},
  {Roseboom}, {Salvato}, {Sanchez-Portal}, {Scott}, {Smail}, {Sturm},
  {Swinbank}, {Tacconi}, {Valtchanov}, {Wang}, \& {Wuyts}}]{magnelli12}
{Magnelli} B. {et~al.}, 2012, \aap, 539, A155

\bibitem[{{Maiolino} {et~al}\mbox{.}(2001){Maiolino}, {Marconi}, {Salvati},
  {Risaliti}, {Severgnini}, {Oliva}, {La Franca}, \& {Vanzi}}]{maiolino01}
{Maiolino} R., {Marconi} A., {Salvati} M., {Risaliti} G., {Severgnini} P.,
  {Oliva} E., {La Franca} F., {Vanzi} L., 2001, \aap, 365, 28

\bibitem[{{Maiolino} \& {Rieke}(1995)}]{maiolino95}
{Maiolino} R., {Rieke} G.~H., 1995, \apj, 454, 95

\bibitem[{{Melbourne} {et~al}\mbox{.}(2012){Melbourne}, {Soifer}, {Desai},
  {Pope}, {Armus}, {Dey}, {Bussmann}, {Jannuzi}, \& {Alberts}}]{melbourne12}
{Melbourne} J. {et~al.}, 2012, \aj, 143, 125

\bibitem[{{Mihos}(1996)}]{mihos96}
{Mihos} C., 1996, in {Encyclopedia of Astronomy and Astrophysics}, {IOP
  Publishing Ltd}

\bibitem[{{Narayanan} {et~al}\mbox{.}(2010){Narayanan}, {Dey}, {Hayward},
  {Cox}, {Bussmann}, {Brodwin}, {Jonsson}, {Hopkins}, {Groves}, {Younger}, \&
  {Hernquist}}]{narayanan10}
{Narayanan} D. {et~al.}, 2010, \mnras, 407, 1701

\bibitem[{{Negrello} {et~al}\mbox{.}(2010){Negrello}, {Hopwood}, {De Zotti},
  {Cooray}, {Verma}, {Bock}, {Frayer}, {Gurwell}, {Omont}, {Neri},
  {Dannerbauer}, {Leeuw}, {Barton}, {Cooke}, {Kim}, {da Cunha}, {Rodighiero},
  {Cox}, {Bonfield}, {Jarvis}, {Serjeant}, {Ivison}, {Dye}, {Aretxaga},
  {Hughes}, {Ibar}, {Bertoldi}, {Valtchanov}, {Eales}, {Dunne}, {Driver},
  {Auld}, {Buttiglione}, {Cava}, {Grady}, {Clements}, {Dariush}, {Fritz},
  {Hill}, {Hornbeck}, {Kelvin}, {Lagache}, {Lopez-Caniego}, {Gonzalez-Nuevo},
  {Maddox}, {Pascale}, {Pohlen}, {Rigby}, {Robotham}, {Simpson}, {Smith},
  {Temi}, {Thompson}, {Woodgate}, {York}, {Aguirre}, {Beelen}, {Blain},
  {Baker}, {Birkinshaw}, {Blundell}, {Bradford}, {Burgarella}, {Danese},
  {Dunlop}, {Fleuren}, {Glenn}, {Harris}, {Kamenetzky}, {Lupu}, {Maddalena},
  {Madore}, {Maloney}, {Matsuhara}, {Michaowski}, {Murphy}, {Naylor}, {Nguyen},
  {Popescu}, {Rawlings}, {Rigopoulou}, {Scott}, {Scott}, {Seibert}, {Smail},
  {Tuffs}, {Vieira}, {van der Werf}, \& {Zmuidzinas}}]{negrello10}
{Negrello} M. {et~al.}, 2010, Science, 330, 800

\bibitem[{{Neugebauer} {et~al}\mbox{.}(1984){Neugebauer}, {Habing}, {van
  Duinen}, {Aumann}, {Baud}, {Beichman}, {Beintema}, {Boggess}, {Clegg}, {de
  Jong}, {Emerson}, {Gautier}, {Gillett}, {Harris}, {Hauser}, {Houck},
  {Jennings}, {Low}, {Marsden}, {Miley}, {Olnon}, {Pottasch}, {Raimond},
  {Rowan-Robinson}, {Soifer}, {Walker}, {Wesselius}, \& {Young}}]{neugebauer84}
{Neugebauer} G. {et~al.}, 1984, \apjl, 278, L1

\bibitem[{{Osterbrock} \& {Shaw}(1988)}]{osterbrock88}
{Osterbrock} D.~E., {Shaw} R.~A., 1988, \apj, 327, 89

\bibitem[{{Piconcelli} {et~al}\mbox{.}(2003){Piconcelli}, {Cappi}, {Bassani},
  {Di Cocco}, \& {Dadina}}]{piconcelli03}
{Piconcelli} E., {Cappi} M., {Bassani} L., {Di Cocco} G., {Dadina} M., 2003,
  \aap, 412, 689

\bibitem[{{Polletta} {et~al}\mbox{.}(2007){Polletta}, {Tajer}, {Maraschi},
  {Trinchieri}, {Lonsdale}, {Chiappetti}, {Andreon}, {Pierre}, {Le F{\`e}vre},
  {Zamorani}, {Maccagni}, {Garcet}, {Surdej}, {Franceschini}, {Alloin},
  {Shupe}, {Surace}, {Fang}, {Rowan-Robinson}, {Smith}, \&
  {Tresse}}]{polletta07}
{Polletta} M. {et~al.}, 2007, \apj, 663, 81

\bibitem[{{Polletta} {et~al}\mbox{.}(2006){Polletta}, {Wilkes}, {Siana},
  {Lonsdale}, {Kilgard}, {Smith}, {Kim}, {Owen}, {Efstathiou}, {Jarrett},
  {Stacey}, {Franceschini}, {Rowan-Robinson}, {Babbedge}, {Berta}, {Fang},
  {Farrah}, {Gonz{\'a}lez-Solares}, {Morrison}, {Surace}, \&
  {Shupe}}]{polletta06}
{Polletta} M.~d.~C. {et~al.}, 2006, \apj, 642, 673

\bibitem[{{Reddy} {et~al}\mbox{.}(2008){Reddy}, {Steidel}, {Pettini},
  {Adelberger}, {Shapley}, {Erb}, \& {Dickinson}}]{reddy08}
{Reddy} N.~A., {Steidel} C.~C., {Pettini} M., {Adelberger} K.~L., {Shapley}
  A.~E., {Erb} D.~K., {Dickinson} M., 2008, \apjs, 175, 48

\bibitem[{{Richards} {et~al}\mbox{.}(2006){Richards}, {Strauss}, {Fan}, {Hall},
  {Jester}, {Schneider}, {Vanden Berk}, {Stoughton}, {Anderson}, {Brunner},
  {Gray}, {Gunn}, {Ivezi{\'c}}, {Kirkland}, {Knapp}, {Loveday}, {Meiksin},
  {Pope}, {Szalay}, {Thakar}, {Yanny}, {York}, {Barentine}, {Brewington},
  {Brinkmann}, {Fukugita}, {Harvanek}, {Kent}, {Kleinman}, {Krzesi{\'n}ski},
  {Long}, {Lupton}, {Nash}, {Neilsen}, {Nitta}, {Schlegel}, \&
  {Snedden}}]{richards06}
{Richards} G.~T. {et~al.}, 2006, \aj, 131, 2766

\bibitem[{{Risaliti} {et~al}\mbox{.}(1999){Risaliti}, {Maiolino}, \&
  {Salvati}}]{risaliti99}
{Risaliti} G., {Maiolino} R., {Salvati} M., 1999, \apj, 522, 157

\bibitem[{{Sanders} \& {Mirabel}(1996)}]{sanders&mirabel96}
{Sanders} D.~B., {Mirabel} I.~F., 1996, \araa, 34, 749

\bibitem[{{Sanders} {et~al}\mbox{.}(1988{\natexlab{a}}){Sanders}, {Soifer},
  {Elias}, {Madore}, {Matthews}, {Neugebauer}, \& {Scoville}}]{sanders88a}
{Sanders} D.~B., {Soifer} B.~T., {Elias} J.~H., {Madore} B.~F., {Matthews} K.,
  {Neugebauer} G., {Scoville} N.~Z., 1988{\natexlab{a}}, \apj, 325, 74

\bibitem[{{Sanders} {et~al}\mbox{.}(1988{\natexlab{b}}){Sanders}, {Soifer},
  {Elias}, {Neugebauer}, \& {Matthews}}]{sanders88b}
{Sanders} D.~B., {Soifer} B.~T., {Elias} J.~H., {Neugebauer} G., {Matthews} K.,
  1988{\natexlab{b}}, \apjl, 328, L35

\bibitem[{{Schweizer}(1998)}]{schweizer98}
{Schweizer} F., 1998, in Saas-Fee Advanced Course 26: Galaxies: Interactions
  and Induced Star Formation, {Kennicutt} Jr. R.~C., {Schweizer} F., {Barnes}
  J.~E., {Friedli} D., {Martinet} L., {Pfenniger} D., eds., p. 105

\bibitem[{{Scott} {et~al}\mbox{.}(2010){Scott}, {Yun}, {Wilson}, {Austermann},
  {Aguilar}, {Aretxaga}, {Ezawa}, {Ferrusca}, {Hatsukade}, {Hughes}, {Iono},
  {Giavalisco}, {Kawabe}, {Kohno}, {Mauskopf}, {Oshima}, {Perera}, {Rand},
  {Tamura}, {Tosaki}, {Velazquez}, {Williams}, \& {Zeballos}}]{scott10}
{Scott} K.~S. {et~al.}, 2010, \mnras, 405, 2260

\bibitem[{{Scott} {et~al}\mbox{.}(2006){Scott}, {Dunlop}, \&
  {Serjeant}}]{scott06}
{Scott} S.~E., {Dunlop} J.~S., {Serjeant} S., 2006, \mnras, 370, 1057

\bibitem[{{Scott} {et~al}\mbox{.}(2002){Scott}, {Fox}, {Dunlop}, {Serjeant},
  {Peacock}, {Ivison}, {Oliver}, {Mann}, {Lawrence}, {Efstathiou},
  {Rowan-Robinson}, {Hughes}, {Archibald}, {Blain}, \& {Longair}}]{scott02}
{Scott} S.~E. {et~al.}, 2002, \mnras, 331, 817

\bibitem[{{Scoville} {et~al}\mbox{.}(2000){Scoville}, {Evans}, {Thompson},
  {Rieke}, {Hines}, {Low}, {Dinshaw}, {Surace}, \& {Armus}}]{scoville00}
{Scoville} N.~Z. {et~al.}, 2000, \aj, 119, 991

\bibitem[{{Smail} {et~al}\mbox{.}(1997){Smail}, {Ivison}, \& {Blain}}]{smail97}
{Smail} I., {Ivison} R.~J., {Blain} A.~W., 1997, \apjl, 490, L5

\bibitem[{{Soifer} {et~al}\mbox{.}(1984){Soifer}, {Rowan-Robinson}, {Houck},
  {de Jong}, {Neugebauer}, {Aumann}, {Beichman}, {Boggess}, {Clegg}, {Emerson},
  {Gillett}, {Habing}, {Hauser}, {Low}, {Miley}, \& {Young}}]{soifer84}
{Soifer} B.~T. {et~al.}, 1984, \apjl, 278, L71

\bibitem[{{Solomon} \& {Vanden Bout}(2005)}]{solomon05}
{Solomon} P.~M., {Vanden Bout} P.~A., 2005, \araa, 43, 677

\bibitem[{{Spoon} {et~al}\mbox{.}(2013){Spoon}, {Farrah}, {Lebouteiller},
  {Gonz{\'a}lez-Alfonso}, {Bernard-Salas}, {Urrutia}, {Rigopoulou},
  {Westmoquette}, {Smith}, {Afonso}, {Pearson}, {Cormier}, {Efstathiou},
  {Borys}, {Verma}, {Etxaluze}, \& {Clements}}]{spoon13}
{Spoon} H.~W.~W. {et~al.}, 2013, \apj, 775, 127

\bibitem[{{Stern} {et~al}\mbox{.}(2014){Stern}, {Lansbury}, {Assef}, {Brandt},
  {Alexander}, {Ballantyne}, {Balokovic}, {Benford}, {Blain}, {Boggs},
  {Bridge}, {Brightman}, {Christensen}, {Comastri}, {Craig}, {Del Moro},
  {Eisenhardt}, {Gandhi}, {Griffith}, {Hailey}, {Harrison}, {Hickox},
  {Jarrett}, {Koss}, {Lake}, {LaMassa}, {Luo}, {Tsai}, {Walton}, {Wright},
  {Wu}, {Yan}, \& {Zhang}}]{stern14}
{Stern} D. {et~al.}, 2014, ArXiv e-prints

\bibitem[{{Tacconi} {et~al}\mbox{.}(2008){Tacconi}, {Genzel}, {Smail}, {Neri},
  {Chapman}, {Ivison}, {Blain}, {Cox}, {Omont}, {Bertoldi}, {Greve},
  {F{\"o}rster Schreiber}, {Genel}, {Lutz}, {Swinbank}, {Shapley}, {Erb},
  {Cimatti}, {Daddi}, \& {Baker}}]{tacconi08}
{Tacconi} L.~J. {et~al.}, 2008, \apj, 680, 246

\bibitem[{{Treister} \& {Urry}(2005)}]{treister05}
{Treister} E., {Urry} C.~M., 2005, \apj, 630, 115

\bibitem[{{Tyler} {et~al}\mbox{.}(2009){Tyler}, {Le Floc'h}, {Rieke}, {Dey},
  {Desai}, {Brand}, {Borys}, {Jannuzi}, {Armus}, {Dole}, {Papovich}, {Brown},
  {Blaylock}, {Higdon}, {Higdon}, {Charmandaris}, {Ashby}, \&
  {Smith}}]{tyler09}
{Tyler} K.~D. {et~al.}, 2009, \apj, 691, 1846

\bibitem[{{Umehata} {et~al}\mbox{.}(2014){Umehata}, {Tamura}, {Kohno},
  {Hatsukade}, {Scott}, {Kubo}, {Yamada}, {Ivison}, {Cybulski}, {Aretxaga},
  {Austermann}, {Hughes}, {Ezawa}, {Hayashino}, {Ikarashi}, {Iono}, {Kawabe},
  {Matsuda}, {Matsuo}, {Nakanishi}, {Oshima}, {Perera}, {Takata}, {Wilson}, \&
  {Yun}}]{umehata14}
{Umehata} H. {et~al.}, 2014, \mnras, 440, 3462

\bibitem[{{Veilleux} {et~al}\mbox{.}(2002){Veilleux}, {Kim}, \&
  {Sanders}}]{veilleux02}
{Veilleux} S., {Kim} D.-C., {Sanders} D.~B., 2002, \apjs, 143, 315

\bibitem[{{Vieira} {et~al}\mbox{.}(2010){Vieira}, {Crawford}, {Switzer}, {Ade},
  {Aird}, {Ashby}, {Benson}, {Bleem}, {Brodwin}, {Carlstrom}, {Chang}, {Cho},
  {Crites}, {de Haan}, {Dobbs}, {Everett}, {George}, {Gladders}, {Hall},
  {Halverson}, {High}, {Holder}, {Holzapfel}, {Hrubes}, {Joy}, {Keisler},
  {Knox}, {Lee}, {Leitch}, {Lueker}, {Marrone}, {McIntyre}, {McMahon}, {Mehl},
  {Meyer}, {Mohr}, {Montroy}, {Padin}, {Plagge}, {Pryke}, {Reichardt}, {Ruhl},
  {Schaffer}, {Shaw}, {Shirokoff}, {Spieler}, {Stalder}, {Staniszewski},
  {Stark}, {Vanderlinde}, {Walsh}, {Williamson}, {Yang}, {Zahn}, \&
  {Zenteno}}]{vieira10}
{Vieira} J.~D. {et~al.}, 2010, \apj, 719, 763

\bibitem[{{Wei{\ss}} {et~al}\mbox{.}(2009){Wei{\ss}}, {Kov{\'a}cs}, {Coppin},
  {Greve}, {Walter}, {Smail}, {Dunlop}, {Knudsen}, {Alexander}, {Bertoldi},
  {Brandt}, {Chapman}, {Cox}, {Dannerbauer}, {De Breuck}, {Gawiser}, {Ivison},
  {Lutz}, {Menten}, {Koekemoer}, {Kreysa}, {Kurczynski}, {Rix}, {Schinnerer},
  \& {van der Werf}}]{weiss09}
{Wei{\ss}} A. {et~al.}, 2009, \apj, 707, 1201

\bibitem[{{Williams} \& {Lewis}(1996)}]{williams96}
{Williams} L.~L.~R., {Lewis} G.~F., 1996, \mnras, 281, L35

\bibitem[{{Wright} {et~al}\mbox{.}(2010){Wright}, {Eisenhardt}, {Mainzer},
  {Ressler}, {Cutri}, {Jarrett}, {Kirkpatrick}, {Padgett}, {McMillan},
  {Skrutskie}, {Stanford}, {Cohen}, {Walker}, {Mather}, {Leisawitz}, {Gautier},
  {McLean}, {Benford}, {Lonsdale}, {Blain}, {Mendez}, {Irace}, {Duval}, {Liu},
  {Royer}, {Heinrichsen}, {Howard}, {Shannon}, {Kendall}, {Walsh}, {Larsen},
  {Cardon}, {Schick}, {Schwalm}, {Abid}, {Fabinsky}, {Naes}, \&
  {Tsai}}]{wright10}
{Wright} E.~L. {et~al.}, 2010, \aj, 140, 1868

\bibitem[{{Wu} {et~al}\mbox{.}(2014){Wu}, {Bussmann}, {Tsai}, {Petric},
  {Blain}, {Eisenhardt}, {Bridge}, {Benford}, {Stern}, {Assef}, {Gelino},
  {Moustakas}, \& {Wright}}]{wu14}
{Wu} J. {et~al.}, 2014, ArXiv 1405.1147

\bibitem[{{Wu} {et~al}\mbox{.}(2012){Wu}, {Tsai}, {Sayers}, {Benford},
  {Bridge}, {Blain}, {Eisenhardt}, {Stern}, {Petty}, {Assef}, {Bussmann},
  {Comerford}, {Cutri}, {Evans}, {Griffith}, {Jarrett}, {Lake}, {Lonsdale},
  {Rho}, {Stanford}, {Weiner}, {Wright}, \& {Yan}}]{wu12}
{Wu} J. {et~al.}, 2012, \apj, 756, 96

\end{thebibliography}

\begin{table*}
\caption{Coordinates and photometry of the 10 Hot DOGs, with 3.4\,$\mu$m, 4.6\,$\mu$m, 12\,$\mu$m and 22\,$\mu$m magnitudes from the AllWISE Source Catalog and 850\,$\mu$m flux densities from SCUBA-2. The top six targets are detected at 850\,$\mu$m, while the bottom four targets have upper limits at 850\,$\mu$m. The targets with WISE upper limits have SNR $<$ 2 and therefore in the AllWISE Source Catalog the magnitudes quoted are 2$\sigma$ upper limits. For redshifts of the targets refer to Eisenhardt et al. (2012, in prep.) and Bridge et al. in prep.}
\begin{tabular}{@{}cccccccccc@{}}
\hline
Source & R.A. & Dec. & 3.4\,$\mu$m & 4.6\,$\mu$m & 12\,$\mu$m & 22\,$\mu$m & 850\,$\mu$m & 850\,$\mu$m / 22\,$\mu$m & Redshift \\
 & (J2000) & (J2000) & (mag) & (mag) & (mag) & (mag) & (mJy) & Ratio & \\
\hline
W0831$+$0140 & 08:31:53.30 & $+$01:40:10.8 & 17.92 $\pm$ 0.28 & 16.10 $\pm$ 0.20 & 10.15 $\pm$ 0.07 & 7.28 $\pm$ 0.12 & 9.3 $\pm$ 2.1 & 0.9 $\pm$ 0.2 & 3.91 \\
W1136$+$4236 & 11:36:34.31 & $+$42:36:02.6 & 18.19 $\pm$ 0.24 & 15.83 $\pm$ 0.11 & 10.62 $\pm$ 0.07 & 7.66 $\pm$ 0.11 & 5.3 $\pm$ 1.7 & 0.7 $\pm$ 0.1 & 2.39 \\
W1603$+$2745 & 16:03:57.40 & $+$27:45:53.3 & $<$ 18.02 & 17.04 $\pm$ 0.34 & 9.89 $\pm$ 0.04 & 7.28 $\pm$ 0.11 & 10.2 $\pm$ 1.8 & 1.0 $\pm$ 0.2 & 2.63 \\
W1835$+$4355 & 18:35:33.73 & $+$43:55:48.7 & 17.44 $\pm$ 0.09 & 15.20 $\pm$ 0.05 & 9.15 $\pm$ 0.03 & 6.19 $\pm$ 0.04 & 8.0 $\pm$ 1.5 & 0.3 $\pm$ 0.1 & 2.30 \\
W2216$+$0723 & 22:16:19.09 & $+$07:23:54.5 & 17.33 $\pm$ 0.16 & 15.59 $\pm$ 0.13 & 9.91 $\pm$ 0.05 & 6.91 $\pm$ 0.09 & 5.5 $\pm$ 1.6 & 0.4 $\pm$ 0.1 & 1.68 \\
W2246$-$0526 & 22:46:07.54 & $-$05:26:35.1 & 17.54 $\pm$ 0.21 & 16.65 $\pm$ 0.37 & 10.27 $\pm$ 0.09 & 6.80 $\pm$ 0.11 & 11.4 $\pm$ 2.1 & 0.7 $\pm$ 0.1 & 4.59 \\
\hline
W1814$+$3412 & 18:14:17.31 & $+$34:12:24.8 & 18.86 $\pm$ 0.44 & 17.61 $\pm$ 0.49 & 10.41 $\pm$ 0.06 & 6.86 $\pm$ 0.07 & 2.0 $\pm$ 1.8 & $<$ 0.4 & 2.45 \\
W2026$+$0716 & 20:26:15.27 & $+$07:16:23.9 & 17.58 $\pm$ 0.21 & 15.69 $\pm$ 0.13 & 10.18 $\pm$ 0.07 & 7.31 $\pm$ 0.11 & 2.0 $\pm$ 1.7 & $<$ 0.6 & 2.54 \\
W2054$+$0207 & 20:54:25.69 & $+$02:07:11.0 & 18.27 $\pm$ 0.32 & 15.32 $\pm$ 0.09 & 9.59 $\pm$ 0.05 & 7.13 $\pm$ 0.09 & 3.3 $\pm$ 1.8 & $<$ 1.1 & 2.52 \\
W2357$+$0328 & 23:57:10.82 & $+$03:28:03.4 & $<$ 18.14 & $<$ 16.61 & 10.09 $\pm$ 0.07 & 6.94 $\pm$ 0.11 & 2.2 $\pm$ 1.9 & $<$ 0.4 & 2.12 \\
\hline
\end{tabular}
\label{fluxes}
\end{table*}

\begin{table*}
\caption{The total IR luminosities (8$\mu \textrm{m}-\textrm{SCUBA2}$) of the 10 Hot DOGs derived by connecting all the WISE and SCUBA-2 data points with power laws and then integrating. The top six have detections at 850\,$\mu$m and the bottom four have 2$\sigma$ upper limits at 850\,$\mu$m. The luminosities are shown in solar luminosities, 3.84 $\times$ 10$^{26}$ W. The W1814$+$3412 template total IR luminosities L$_{8-1000\mu \textrm{m}}$ are found by using the Blain et al. (in prep.) W1814$+$3412 template with the Hot DOGs' SCUBA-2 data.
The total IR luminosities (8-1000\,$\mu$m) of two targets also observed by \citet{wu12} are 4.0 $\times$ 10$^{13}$\,L$_{\odot}$ for W1814$+$3412, and 6.5 $\times$ 10$^{13}$\,L$_{\odot}$ for W1835$+$4355, are consistent with the luminosities in the table below.}
\begin{tabular}{@{}ccc@{}}
\hline
Source & Total IR Luminosities (8$\mu \textrm{m}-\textrm{SCUBA2}$) & W1814$+$3412 Template \\
 & (L$_\odot$) & Total IR Luminosities (8-1000\,$\mu$m) \\
 & & (L$_\odot$) \\
\hline
W0831$+$0140 & 8.7 $\pm$ 1.8 $\times$ 10$^{13}$ & 3.6 $\pm$ 1.6 $\times$ 10$^{14}$\\
W1136$+$4236 & 1.5 $\pm$ 4.6 $\times$ 10$^{13}$ & 6.2 $\pm$ 3.8 $\times$ 10$^{13}$\\
W1603$+$2745 & 3.1 $\pm$ 0.7 $\times$ 10$^{13}$ & 1.5 $\pm$ 0.5 $\times$ 10$^{14}$\\
W1835$+$4355 & 4.3 $\pm$ 4.1 $\times$ 10$^{13}$ & 8.5 $\pm$ 3.4 $\times$ 10$^{13}$\\
W2216$+$0723 & 1.0 $\pm$ 1.8 $\times$ 10$^{13}$ & 2.7 $\pm$ 1.6 $\times$ 10$^{13}$\\
W2246$-$0526 & 1.3 $\pm$ 2.9 $\times$ 10$^{14}$ & 6.4 $\pm$ 2.4 $\times$ 10$^{14}$\\
\hline
W1814$+$3412 & $<$ 2.5 $\times$ 10$^{13}$ & $<$ 7.0 $\times$ 10$^{13}$\\
W2026$+$0716 & $<$ 2.1 $\times$ 10$^{13}$ & $<$ 7.3 $\times$ 10$^{13}$\\
W2054$+$0207 & $<$ 2.9 $\times$ 10$^{13}$ & $<$ 9.2 $\times$ 10$^{13}$\\
W2357$+$0328 & $<$ 1.9 $\times$ 10$^{13}$ & $<$ 5.3 $\times$ 10$^{13}$\\
\hline
\end{tabular}
\label{luminosities}
\end{table*}

\begin{table*}
\caption{Number of serendipitous sources in each of the 10 maps at greater than 3$\sigma$ and 4$\sigma$ significance, and the number of negative peaks at greater than 3$\sigma$ significance. There were no negative peaks at greater than 4$\sigma$ in the 10 maps.}
\begin{tabular}{@{}cccc@{}}
\hline
Source & Number of & Number of & Number of \\
 & Serendipitous Sources & Serendipitous Sources & Negative Peaks \\
 & at greater than 3$\sigma$ & at greater than 4$\sigma$ & at greater than 3$\sigma$ \\
\hline
W0831$+$0140 &  3 & 0 & 0\\
W1136$+$4236 &  1 & 0 & 0\\
W1603$+$2745 &  1 & 0 & 0\\
W1814$+$3412 &  1 & 0 & 1\\
W1835$+$4355 &  1 & 0 & 0\\
W2026$+$0716 &  2 & 1 & 1\\
W2054$+$0207 &  3 & 0 & 1\\
W2216$+$0723 &  2 & 0 & 0\\
W2246$-$0526 &  1 & 0 & 1\\
W2357$+$0328 &  2 & 0 & 0\\
\hline
\end{tabular}
\label{number}
\end{table*}

\label{lastpage}

\end{document}